\documentclass
[twocolumn,amsmath,amssymb,amsfonts,prb,superscriptaddress]{revtex4}%
\usepackage{graphicx}
\usepackage{dcolumn}
\usepackage{bm}
\usepackage[bookmarks=false]{hyperref}
\usepackage{hypernat}
\allowdisplaybreaks[1]

\begin{document}
\title{Bosonization approach to the mixed-valence two-channel Kondo problem}
\author{A.~Iucci}
\affiliation{DPMC-MaNEP, University of Geneva, 24 Quai Ernest Ansermet, CH-1211 Geneva 4, Switzerland}
\author{C.~J.~Bolech}
\affiliation{DPMC-MaNEP, University of Geneva, 24 Quai Ernest Ansermet, CH-1211 Geneva 4, Switzerland}
\affiliation{Physics \& Astronomy Department, Rice University, 6100 Main Street, Houston, TX-77005, USA}
\date{September 21$^\text{st}$, 2007}

\begin{abstract}
We present in detail the bosonization-refermionization solution of the
anisotropic version of the two-channel Anderson model at a particular
manifold in the space of parameters of the theory, where we establish
an equivalence with a Fermi-Majorana biresonant-level model. The
correspondence is rigorously proved by explicitly constructing the new
fermionic fields and Klein factors in terms of the original ones and
showing that the commutation properties between original and new Klein
factors are of semionic type. We also demonstrate that the fixed
points associated with the solvable manifold are renormalization-group
stable and generic, and therefore representative of the physics of the
original model. The simplicity of the solution found allows for the
computation of the full set of thermodynamic quantities. In particular,
we compute the entropy, occupation, and magnetization of the impurity
as functions of temperature, and identify the different physical
energy scales. In the absence of external fields, two energy scales appear
and, as the temperature goes to zero, a nontrivial residual entropy
indicates that the model approaches a universal line of fixed points
of non-Fermi-liquid type. An external field, even if small, introduces
a third energy scale and causes the quenching of the impurity entropy
to zero, taking the system to a corresponding Fermi-liquid fixed
point.

\end{abstract}

\maketitle

\section{Introduction}

The ideas that motivated the introduction of the two-channel
Anderson model started with an attempt to describe the
non-Fermi-liquid physics of the \textrm{UBe}$_{13}$ compound and
other \textrm{U}-based heavy fermions.\cite{cox1987} The model can
be thought of as a mixed-valence model that, in certain regimes,
has the two-channel Kondo model \cite{Nozieres1980} as its
low-energy theory. In that regard, the relation between the two
models is as in the case of the single-channel Kondo and Anderson
models; the Anderson model captures the local-moment physics and
provides a physical mechanism for moment formation, while at the
same time describes also higher temperature and degenerate-level
regimes for which mixed valence prevails and there are no
localized moments. The study of models that capture such regimes
is important, since there exist a growing number of compounds
that are believed to display mixed valence [besides the large
number of \textrm{U}-based compounds, recently, a number of new
\textrm{Pr}-based heavy fermions were proposed as experimental
candidates for the realization of quadrupolar-Kondo ground states;
for instance, \textrm{La}-doped
\textrm{PrPb}$_{3}$ (Ref.~\onlinecite{kawae2006})]. Hence, improving our
comprehension of the Anderson model physics is important for the
phenomenological description of all these compounds.\cite{cox1998}
The peculiarity of \textrm{U} or \textrm{Pr} ions compared to
the ions in other typical heavy-Fermion compounds is that, in a
cubic crystal field, the $5f^{2}$ configuration of \textrm{U}$^{+4}$
ions or the $4f^{2}$ configuration of \textrm{Pr}$^{+3}$ are
projected into a non-Kramers $\Gamma_{3}$ doublet with a
quadrupolar moment. Fluctuations out of this state will hybridize
with Kramers-degenerate states and provide a competition between
magnetic and quadrupolar (\textit{i.e.},~flavor) moments. A minimal
model in the case of the uranium compounds, that takes into
account spin-orbit and crystal-field effects, leads to modeling
those two states with $\Gamma_{3}$ (flavor) and $\Gamma_{6}$
(spin) doublets that hybridize with $\Gamma_{8}$ conduction
electrons and give rise to a description in terms of the
two-channel Anderson model.\cite{cox1998} The situation is similar
in the case of praseodymium compounds.\cite{hotta2005,hotta2006}

For the case of \textrm{UBe}$_{13}$, it was originally speculated that
the $\Gamma_3$ doublet would constitute the lowest-energy ionic
configuration. This opened the possibility of a
\textit{quadrupolar route} to two-channel Kondo physics, in which a
local quadrupolar moment would be screened by conduction electrons
whose spin degree of freedom would provide the two degenerate
channels.\cite{cox1987} Nonlinear susceptibility measurements
disfavored such scenario,\cite{ramirez1994} however, and indicated
instead the possibility of a mixed-valence state.\cite{aliev1995} The
latter was also supported by de Haas--van Alphen measurements of
uranium-doped samples of \textrm{ThBe}$_{13}$.\cite{harrison2001} The
new scenario necessitates the study of the \textit{full} two-channel
Anderson model away from its local-moment regimes, where a
low-energy Kondo-Hamiltonian description is not available. Subsequent
theoretical studies of the model firmly established the persistence of
local non-Fermi-liquid physics over its whole parameter regime,
including at mixed valence.\cite{schiller1998,koga1999,bolech2002} For
a contrast, one could call this the \textit{mixed-valence route} to
two-channel Kondo physics.  In contradistinction, the quadrupolar
route, --practically dismissed for uranium compounds--, recently
acquired new relevance in the context of certain praseodymium
compounds. Measurements on
\textrm{PrPb}$_3$,\cite{kawae2006,onimaru2005}
\textrm{PrOs}$_4$\textrm{Sb}$_{12}$,\cite{bauer2002,bauer2006} and other
\textrm{Pr}-based skutterudite compounds are consistent with the
possibility of nonmagnetic ionic ground states and either onsite or offsite
(antiferro)-quadrupolar fluctuations. For a summarized account of some
of these experimental results, put in the context of Kondo physics,
see, for instance, Ref.~\onlinecite{maple2005}.

In parallel, the Kondo effect became an important subject of study in
the field of mesoscopics and there are many proposals and experiments
for realizing two-channel Kondo systems using tunable setups such as
quantum dots.\cite{berman1999,oreg2003,shah2003,kakashvili2007,potok2007}
Experimental realizations based on the two-channel Anderson model may
be more robust and allow for the observation of Kondo physics at
higher temperatures, since the Kondo scale increases exponentially as
the system goes into the mixed-valence regime; the Kondo physics
should still be observable in quantities like the charge fluctuations
encoded in the measurements of capacitance
lineshapes.\cite{bolech2005b}

Since the original work of Kondo,\cite{kondo1964} the study of
quantum impurity models evolved rapidly. It quickly became evident
that other methods going beyond perturbation theory were required in
order to access the low-temperature physics of these models. The
most conspicuous achievements in this front were the numerical
renormalization group developed by Wilson \cite{wilson1975} and
the Bethe ansatz solution obtained independently by Andrei and
Wiegmann.\cite{andrei1980,wiegmann1980} Another important
development was the identification by Toulouse
\cite{toulouse1969,schlottmann1978} of a solvable line on which an
anisotropic version of the model is mapped into a fermionic
resonant-level model.\cite{vigman1978} These non-perturbative
techniques are required in order to study the crossover regimes
and allow the unambiguous identification of the strongly coupled
fixed point of the quantum impurities. Numerical renormalization
group and Bethe ansatz were both applied successfully to
Hamiltonians that incorporate valence fluctuation
physics,\cite{Hewson} but mappings like the one discovered by
Toulouse remained mostly restricted to
exchange models with fixed valence (cf.~Ref.~%
%TCIMACRO{\TeXButton{kotliar1996}{\onlinecite{kotliar1996}}}%
%BeginExpansion
\onlinecite{kotliar1996}%
%EndExpansion
). In the present work, we address this missing link.\cite{bolech2006a}

As compared to other non-perturbative techniques applied to quantum
impurity problems,\cite{afl1983,wilson1975,affleck1995} bosonization
--or Coulomb gas-- based mappings are complementary and especially
valuable in that they provide us with simple alternative ways of
visualizing the physics,\cite{Giamarchi} and, in particular, the
different crossovers. In this paper, we examine a mapping between
the anisotropic two-channel Anderson impurity model and a particularly
simple biresonant-level Hamiltonian. Our work generalizes the results
of the Emery-Kivelson mapping for the two-channel Kondo Hamiltonian
\cite{emery1992,fabrizio1995,schofield1997,ye1997,vondelft1998} to
a more involved and descriptive model that contains, as well, the
physics of charge fluctuations. After the mapping, it becomes
simpler to identify the different crossover energy scales of the
problem and to infer the existence of a line of non-Fermi-liquid
fixed points that governs the low-energy
physics.\cite{johannesson2003,johannesson2005} Using a
renormalization group analysis, we establish the generic nature of
our low-temperature results and their relevance for the original
two-channel Anderson model. We show how to calculate all the
dynamical and thermodynamical quantities of interest pertaining to
the impurity over the full range of parameters and connect
explicitly all the different temperature regimes of the system. We
rederive, in a compellingly compact language, all the results
obtained previously for the model using a variety of other
non-perturbative techniques,
\cite{schiller1998,bolech2002,anders2005,bolech2005a} and obtain a
number of additional results for the situation when external fields are
present and the character of the infrared fixed points is
modified.

The rest of the article is organized as follows. In
Sec.~\ref{sec:model}, we introduce the anisotropic two-channel
Anderson model, including possible external fields acting on the
impurity. In Sec.~\ref{sec:bosonization}, we establish the mapping,
for a set of couplings belonging to a certain manifold, onto a
non-interacting Fermi-Majorana biresonant-level model. The mapping is
carried out in a bosonized language and particular emphasis is put on
the careful treatment of Klein factors. In
Sec.~\ref{sec:stability}, we perform a renormalization group
analysis of the stability of the fixed points contained in the soluble
manifold. In Sec.~\ref{sec:thermodynamics}, we identify the three
crossover scales of the model and compute thermodynamic quantities in
the entire temperature range and for arbitrary values of the external
fields.  Finally, in Sec.~\ref{sec:conclusion}, we provide a
summary of our conclusions and an outlook of the applications of the
mapping to other problems involving two-channel Anderson model
physics.

\section{The anisotropic two-channel Anderson model}

\label{sec:model}

We shall consider a generalized version of the two-channel Anderson model to
which we add terms that break the rotation invariance in spin and flavor
spaces. This is akin to the standard practice in the case of the
single-channel Kondo model of considering an exchange coupling constant that
acquires a different value along the $z$-axis. We denote the Hamiltonian as
\begin{equation}
H=H_{\text{host}}+H_{\text{imp}}+H_{\text{hyb}}+H_{\text{field}}%
+H_{3}~\text{.}%
\end{equation}
The first term describes the dynamics of the band electrons ($\psi
_{\alpha\sigma}^{\dagger}$ where $\sigma=\uparrow,\downarrow$ and $\alpha=+,-$
correspond to the spin and flavor degrees of freedom, respectively) in the
standard approximation of a linearized band dispersion around the Fermi level
fixed by the normal order prescription,%
\begin{equation}
H_{\text{host}}=\sum_{\alpha\sigma}\int dx~:\psi_{\alpha\sigma}^{\dagger
}\left(  x\right)  \left(  -iv_{\text{F}}\partial_{x}\right)  \psi
_{\alpha\sigma}\left(  x\right)  :~\text{.}%
\end{equation}
The second and third terms contain the isolated-impurity contribution and the
band-impurity hybridization, respectively,%
\begin{subequations}
\begin{align}
H_{\text{imp}}  &  =\varepsilon_{s}\sum_{\sigma}X_{\sigma\sigma}%
+\varepsilon_{f}\sum_{\alpha}X_{\bar{\alpha}\bar{\alpha}}~\text{,}\\
H_{\text{hyb}}  &  =V\sum_{\alpha\sigma}\left[  X_{\sigma\bar{\alpha}}%
\psi_{\alpha\sigma}\left(  0\right)  +\psi_{\alpha\sigma}^{\dagger}\left(
0\right)  X_{\bar{\alpha}\sigma}\right]  ~\text{.}%
\end{align}
Here, we have used Hubbard-operator notation to describe the impurity
degrees of freedom ($X_{ab}=\left\vert a\right\rangle \left\langle
b\right\vert $ where $a,b=\sigma,\bar{\alpha}$ and the bar stands for
the complex-conjugate representation). As compared to
slave-operator notation, the use of Hubbard operators automatically
restricts the Hilbert space of the impurity to the physical
one.\cite{bolech2005a} These first three terms constitute the standard
two-channel Anderson model. The fourth term describes the coupling to
external fields:%
\end{subequations}
\begin{equation}
H_{\text{field}}=h_{s}(X_{\downarrow\downarrow}-X_{\uparrow\uparrow}%
)+h_{f}\left(  X_{\bar{+}\bar{+}}-X_{\bar{-}\bar{-}}\right)  ~\text{,}%
\end{equation}
where, for the sake of generality, we included two fields: one coupling to the
impurity spin and the other one coupling to its flavor. Finally, the fifth
term provides the generalized anisotropy and can be written as a sum over
charge ($c$), spin($s$), flavor ($f$), and spin-flavor ($s\!f$) sectors (as
will be seen below, these sectors arise naturally after bosonizing the model),%
\begin{equation}
H_{3}=\sum_{\nu=c,s,f,s\!f}H_{3}^{\nu}~\text{,}%
\end{equation}
with each term taking the form of a density-density interaction between the
different impurity densities with the corresponding ones from the band,%
\begin{equation}
H_{3}^{\nu}=J_{\nu}^{3}X_{\nu}\rho_{\nu}~\text{,}%
\end{equation}
where $\rho_{\nu}\equiv\rho_{\nu}\left(  x=0\right)  $,%
\begin{equation}
\rho_{\nu}\left(  x\right)  =\sum_{\alpha\sigma\alpha^{\prime}\sigma^{\prime}%
}:\psi_{\alpha\sigma}^{\dagger}\left(  x\right)  \Upsilon_{\alpha\sigma
,\alpha^{\prime}\sigma^{\prime}}^{\nu}\psi_{\alpha^{\prime}\sigma^{\prime}%
}\left(  x\right)  :
\end{equation}
and%
\begin{subequations}
\begin{align}
\Upsilon_{\alpha\sigma,\alpha^{\prime}\sigma^{\prime}}^{c}  &  =\delta
_{\alpha\alpha^{\prime}}\delta_{\sigma\sigma^{\prime}}~\text{,} &
\Upsilon_{\alpha\sigma,\alpha^{\prime}\sigma^{\prime}}^{s}  &  =\delta
_{\alpha\alpha^{\prime}}\tau^{3}{}_{\sigma\sigma^{\prime}}~\text{,}\\
\Upsilon_{\alpha\sigma,\alpha^{\prime}\sigma^{\prime}}^{f}  &  =\tau^{3}%
{}_{\alpha\alpha^{\prime}}\delta_{\sigma\sigma^{\prime}}~\text{,} &
\Upsilon_{\alpha\sigma,\alpha^{\prime}\sigma^{\prime}}^{s\!f}  &  =\tau^{3}%
{}_{\alpha\alpha^{\prime}}\tau^{3}{}_{\sigma\sigma^{\prime}}~\text{,}%
\end{align}
(here $\tau^{3}$ is the third Pauli matrix). The impurity densities involved
are%
\end{subequations}
\begin{subequations}
\begin{gather}
X_{s}=\sum_{\sigma}\sigma X_{\sigma\sigma}~\text{,}\\
X_{f}=\sum_{\alpha}\alpha X_{\bar{\alpha}\bar{\alpha}}~\text{,}\\
X_{c}=X_{s\!f}=\sum_{\sigma}X_{\sigma\sigma}-\sum_{\alpha}X_{\bar{\alpha}%
\bar{\alpha}}~\text{.}%
\end{gather}
Since the impurity Hilbert space contains four states, we can define only
three independent density-like operators apart from the identity; we find it
convenient to define the spin-flavor density to coincide with the charge one.

\section{Bosonization based mapping}
\label{sec:bosonization}

Bosonization is a well known technique that renders more accessible the study
of 1+1 dimensional models.\cite{haldane1981,vondelft1998a,Gogolin,Giamarchi}
When combined with refermionization procedures, the bosonic language can be
used to find nontrivial mappings between different fermionic models. The idea
is that, many times, a mapping that results highly nonlinear in the fermionic
language can be written as a simple canonical transformation in terms of the bosons.

\subsection{Bosonization}

We shall employ the standard bosonic representation of the fermionic fields,
\end{subequations}
\begin{equation}
\psi_{\alpha\sigma}\left(  x\right)  =\frac{1}{\sqrt{2\pi a}}F_{\alpha\sigma
}e^{-i\phi_{\alpha\sigma}\left(  x\right)}~\text{,}%
\end{equation}
in which $a$ is a regulator that plays the role of an inverse
bandwidth and $\phi_{\alpha\sigma}$ are bosonic fields that describe
the particle-hole excitations around the Fermi sea. Finally,
$F_{\alpha\sigma}$ are the so called Klein factors, responsible for
recovering the correct anticommutation relations among different
fermionic species and necessary for describing processes in which the
number of fermions changes. They act as ladder operators in the
fermionic Hilbert space\footnote{More explicitly,
$(F_{\alpha\sigma}^{[\dagger]})^n$ indicates a deviation by $n$
particles below [above] from the reference level adopted when normal
ordering the Hamiltonian. In the two-channel Anderson model, these
deviations are generated by the hybridization term in the Hamiltonian
and are restricted by the algebra of the impurity to be less or equal
to 1. For the discussion to come, notice that in the new basis
the spin-flavor sector will not be so restricted, however, which will
explain its ability to hybridize with a Majorana degree of freedom
from the impurity.} and commute with the $\phi$'s.
Additionally, they satisfy the following algebra:%
\begin{subequations}
\begin{align}
F_{\alpha\sigma}^{\dagger}F_{\alpha\sigma}  &  =F_{\alpha\sigma}%
F_{\alpha\sigma}^{\dagger}=1~\text{,}\label{eq:klein_first}\\
F_{\alpha\sigma}^{\dagger}F_{\alpha^{\prime}\sigma^{\prime}}  &
=-F_{\alpha^{\prime}\sigma^{\prime}}F_{\alpha\sigma}^{\dagger}\qquad
\text{for}~\left(  \alpha\sigma\right)  \neq\left(  \alpha^{\prime}%
\sigma^{\prime}\right)  \text{,}\\
F_{\alpha\sigma}F_{\alpha^{\prime}\sigma^{\prime}}  &  =-F_{\alpha^{\prime
}\sigma^{\prime}}F_{\alpha\sigma}\qquad\text{for}~\left(  \alpha\sigma\right)
\neq\left(  \alpha^{\prime}\sigma^{\prime}\right)
\end{align}
and they obey the following (anti)commutation relations with the impurity
Hubbard operators:%
\end{subequations}
\begin{subequations}
\begin{align}
\left[  F_{\alpha\sigma},X_{\bar{\alpha}^{\prime}\bar{\alpha}^{\prime\prime}%
}\right]   &  =\left[  F_{\alpha\sigma},X_{\sigma^{\prime}\sigma^{\prime
\prime}}\right]  =0~\text{,}\\
\left\{  F_{\alpha\sigma},X_{\sigma^{\prime}\bar{\alpha}^{\prime}}\right\}
&  =\left\{  F_{\alpha\sigma},X_{\bar{\alpha}^{\prime}\sigma^{\prime}%
}\right\}  =0~\text{.} \label{eq:klein_last}%
\end{align}

In terms of the bosons, the Hamiltonian for the band takes the form%
\end{subequations}
\begin{equation}
H_{\text{host}}=\frac{v_{\text{F}}}{4\pi}\sum_{\alpha\sigma}\int dx:\left(
\partial_{x}\phi_{\alpha\sigma}\right)  ^{2}:~\text{.}%
\end{equation}
Following Emery and Kivelson, it is natural to introduce a rotated basis for
the bosons ($\nu=c$, $s$, $f$, $s\!f$)%
\begin{align}
\phi_{\nu}  &  =\frac{1}{2}\sum_{\alpha\sigma}\Upsilon_{\alpha\sigma
,\alpha\sigma}^{\nu}\phi_{\alpha\sigma}~\text{, or}\\
\phi_{\alpha\sigma}  &  =\frac{1}{2}\left(  \phi_{c}+\sigma\phi_{s}+\alpha
\phi_{f}+\alpha\sigma\phi_{s\!f}\right)  ~\text{,}%
\end{align}
where $\sigma,\alpha=\pm$ when entering as multiplying factors. The form of
$H_{\text{host}}$ remains the same in the new basis:
\begin{equation}
H_{\text{host}}=\sum_{\nu=c,s,f,s\!f}H_{0}^{\nu}\equiv\frac{v_{\text{F}}}%
{4\pi}\sum_{\nu=c,s,f,s\!f}\int dx:\left(  \partial_{x}\phi_{\nu}\right)
^{2}:~\text{.}%
\end{equation}
Notice that the Klein factors disappeared from the bosonized version of the
first term, but they will enter explicitly in the hybridization term:%
\begin{equation}
H_{\text{hyb}}=\frac{V}{\sqrt{2\pi a}}\sum_{\alpha\sigma}X_{\sigma\bar{\alpha
}}F_{\alpha\sigma}e^{-i\left(  \phi_{c}+\sigma\phi_{s}+\alpha\phi_{f}%
+\alpha\sigma\phi_{s\!f}\right)  /2}+\text{h.c.}%
\end{equation}
Last, the terms involving only impurity operators stay unchanged and the terms
involving exchange are bosonized according to the standard prescription for
densities:
\begin{equation}
H_{3}^{\nu}=J_{\nu}X_{\nu}\rho_{\nu}=-\frac{J_{\nu}}{\pi}X_{\nu}\partial
_{x}\phi_{\nu}(0)~\text{.}%
\end{equation}

The difficulty for studying this model is contained in the highly nontrivial
form of the hybridization with the impurity. Therefore, the strategy is to
look for a canonical transformation that simplifies this term. We define the
following generic transformation $U=U_{c}U_{s}U_{f}U_{s\!f}$ with
\begin{equation}
U_{\nu}=e^{i\gamma_{\nu}\phi_{\nu}(0)X_{\nu}}~\text{.}%
\end{equation}
For transforming $H_{\text{hyb}}$, we first notice that $U$ commutes with the
vertex operators and the Klein factors. Thus, we only need to compute
expressions of the form $U_{\nu}X_{\sigma\bar{\alpha}}U_{\nu}^{\dagger}$.
Using
\begin{subequations}
\begin{align}
\left[  X_{\sigma\bar{\alpha}},X_{s}\right]   &  =-\sigma X_{\sigma\bar
{\alpha}}~\text{,} & \left[  X_{\sigma\sigma},X_{\nu}\right]   &
=0~\text{,}\\
\left[  X_{\sigma\bar{\alpha}},X_{c}\right]   &  =-2X_{\sigma\bar{\alpha}%
}~\text{,} & \left[  X_{\bar{\alpha}\bar{\alpha}},X_{\nu}\right]   &
=0~\text{,}\\
\left[  X_{\sigma\bar{\alpha}},X_{f}\right]   &  =\alpha X_{\sigma\bar{\alpha
}}~\text{,} &
\end{align}
(note that the two identities on the right imply that $H_{\text{imp}}$ and
$H_{\text{field}}$ are not affected by the transformation), we obtain:
\end{subequations}
\begin{subequations}
\begin{align}
U_{s}X_{\sigma\bar{\alpha}}U_{s}^{\dagger}  &  =X_{\sigma\bar{\alpha}%
}e^{\sigma i\gamma_{s}\phi_{s}}~\text{,}\\
U_{f}X_{\sigma\bar{\alpha}}U_{f}^{\dagger}  &  =X_{\sigma\bar{\alpha}%
}e^{-\alpha i\gamma_{f}\phi_{f}}~\text{,}\\
U_{c}X_{\sigma\bar{\alpha}}U_{c}^{\dagger}  &  =X_{\sigma\bar{\alpha}%
}e^{2i\gamma_{c}\phi_{c}}~\text{,}\\
U_{s\!f}X_{\sigma\bar{\alpha}}U_{s\!f}^{\dagger}  &  =X_{\sigma\bar{\alpha}%
}e^{2i\gamma_{s\!f}\phi_{s\!f}}~\text{.}%
\end{align}
Therefore, we can transform the hybridization Hamiltonian as $\tilde
{H}_{\text{hyb}}=UH_{\text{hyb}}U^{\dagger}$, with%
\end{subequations}
\begin{align}
\tilde{H}_{\text{hyb}}  &  =\frac{V}{\sqrt{2\pi a}}\sum_{\alpha\sigma}\left[
X_{\sigma\bar{\alpha}}F_{\alpha\sigma}e^{-i(\frac{1}{2}-2\gamma_{c})\phi_{c}%
}e^{-i\alpha(\frac{1}{2}+\gamma_{f})\phi_{f}}\right. \nonumber\\
&  \times\left.  e^{-i\sigma(\frac{1}{2}-\gamma_{s})\phi_{s}}e^{-i(\frac
{\alpha\sigma}{2}-2\gamma_{s\!f})\phi_{s\!f}}+\text{h.c.}\right]  ~\text{.}%
\end{align}
Next, we use the freedom of choosing the particular transformation
that will simplify the form that this term takes. Such a choice is
given by taking $\gamma_{c}=1/4$, $\gamma_{f}=-1/2$ and
$\gamma_{s}=1/2$, and thus decoupling the impurity from the
corresponding sectors in the band. The transformed expression is%
\begin{multline}
\tilde{H}_{\text{hyb}}=\frac{V}{\sqrt{2\pi a}}\left[  \left(  X_{\uparrow
+}F_{\uparrow+}+X_{\downarrow-}F_{\downarrow-}\right)  e^{-i(\frac{1}%
{2}-2\gamma_{s\!f})\phi_{s\!f}}\right. \\
+\left.  \left(  X_{\uparrow-}F_{\uparrow-}+X_{\downarrow+}F_{\downarrow
+}\right)  e^{-i(-\frac{1}{2}-2\gamma_{s\!f})\phi_{s\!f}}+\text{h.c.}\right]
~\text{,}%
\end{multline}
(where we have omitted the bar over the values taken by the $\bar{\alpha}$
subindex). We can further simplify this term by choosing $\gamma_{s\!f}%
=\pm1/4$. The two choices are equivalent and we choose $\gamma_{s\!f}=1/4$;
in this way, we kill the vertex in the first term while giving the second one
fermionic dimensions:
\begin{multline}
\tilde{H}_{\text{hyb}}=\frac{V}{\sqrt{2\pi a}}\left[  \left(  X_{\uparrow
+}F_{\uparrow+}+X_{\downarrow-}F_{\downarrow-}\right)  \right. \\
+\left.  \left(  X_{\uparrow-}F_{\uparrow-}+X_{\downarrow+}F_{\downarrow
+}\right)  e^{i\phi_{s\!f}}+\text{h.c.}\right]  ~\text{.}%
\end{multline}
The spin-flavor band and the impurity remain coupled, but the form of their
hybridization is now much simpler. Before studying this term further, let us
indicate what happens to the remaining terms in the Hamiltonian upon the same
transformation:%
\begin{multline}
U\left(  H_{\text{host}}+H_{3}\right)  U^{\dagger}=H_{\text{host}}\\
+\frac{1}{\pi}\sum_{\nu}(\pi v_{F}\gamma_{\nu}-J_{\nu})\partial_{x}\phi_{\nu
}(0)X_{\nu}~\text{.}%
\end{multline}
We define for future reference the couplings%
\begin{equation}
\lambda_{\nu}=\pi v_{\text{F}}\gamma_{\nu}-J_{\nu}~\text{,}%
\end{equation}
which can be set to be zero by tuning the values of
$J_{\nu}\rightarrow\pi v_{\text{F}}\gamma_{\nu}$; thus rendering the
model particularly simple, as it will be seen below.

\subsection{Refermionization}

Inspired by the classic results by Toulouse and by Emery and Kivelson
in which they map the single- and the two-channel Kondo models,
respectively, into different types of resonant-level models and find
particular values of the anisotropy for which those resonant-level
models are non-interacting and therefore exactly solvable, we find it
compelling to seek a similar mapping for the two-channel Anderson
model. However, while in the Kondo model the interactions between the
band electrons and the impurity are exchange terms that involve only
fermionic bilinears, that is not the case for the Anderson model in
which the hybridization term allows for the flow of charge between the
band and the impurity.  This is an important difference that gives
rise to a much richer parameter regime, in the case of the Anderson
model, that includes not only local-moment phases but also mixed-valence
ones. From the technical point of view, it becomes important
to keep track of the charge transfer processes while bosonizing, which
requires a proper treatment of the Klein factors.

\subsubsection{New fermions}

In order to find a mapping to a new fermionic model, we need to
refermionize the version that was already simplified by the
cal transformation just discussed. The first step is to
introduce a new set of Klein factors that allow us to define new
fermionic operators corresponding to the physical
sectors ($c$, $s$, $f$, $s\!f$):%

\begin{equation}
\psi_{\nu}=\frac{1}{\sqrt{2\pi a}}F_{\nu}e^{-i\phi_{\nu}}~\text{.}%
\end{equation}
The new Klein factors are defined to be ladder operators in the physical
Hilbert space for the band and to obey the usual algebra given in
Eqs.~(\ref{eq:klein_first})--(\ref{eq:klein_last}).

Inspecting $\tilde{H}_{\text{hyb}}$, we are led to define the new impurity
operators:
\begin{subequations}
\begin{align}
d  &  =F_{s\!f}\left(  X_{\uparrow-}F_{\uparrow-}+X_{\downarrow+}%
F_{\downarrow+}\right)  ~\text{,}\\
f  &  =X_{-+}F_{f}+X_{\uparrow\downarrow}F_{s}~\text{.}%
\end{align}
They allow us to rewrite the hybridization term as%
\end{subequations}
\begin{multline}
\tilde{H}_{\text{hyb}}=\frac{V}{\sqrt{2\pi a}}\left[  \left(  f^{\dagger
}+f\right)  d+d^{\dagger}\left(  f^{\dagger}+f\right)  \right] \\
+V\left(  \psi_{s\!f}^{\dagger}d+d^{\dagger}\psi_{s\!f}\right)  ~\text{.}
\label{eq:Hhyb_intermediate}%
\end{multline}
The new form of the Hamiltonian looks compellingly simple.
However, we will find that the new operators are not fermions. It
can be seen that they satisfy the following relations:
\begin{subequations}
\begin{align}
d^{2}  &  =d^{\dagger2}=0~\text{,}\\
dd^{\dagger}  &  =X_{\uparrow\uparrow}+X_{\downarrow\downarrow}~\text{,}\\
d^{\dagger}d  &  =X_{++}+X_{--}%
\end{align}
and
\end{subequations}
\begin{subequations}
\begin{align}
f^{2}  &  =f^{\dagger2}=0~\text{,}\\
ff^{\dagger}  &  =X_{--}+X_{\uparrow\uparrow}~\text{,}\\
f^{\dagger}f  &  =X_{++}+X_{\downarrow\downarrow}~\text{.}%
\end{align}
Using the completeness relation for the impurity Hilbert space
($X_{\uparrow\uparrow}+X_{\downarrow\downarrow}+X_{++}+X_{--}=1$),
it is clear that both $d$ and $f$ are \emph{self-fermions}
(\textit{i.e.}, each of them independently obeys fermionic
anticommutation relations with itself). We still need to verify
the mixed commutation relations. It is important to stress that
the relations written above can be deduced independently of any
relation between old and new Klein factors. That is not the case
for the relations mixing $d$ and $f$.

\subsubsection{Klein factor relations}

Following the authors of Refs.~\onlinecite{vondelft1998} and
\onlinecite{zarand2000}, we have the freedom to make the following
four identifications between the sets of old and new Klein factors:
\end{subequations}
\begin{subequations}
\begin{align}
F_{s\!f}^{\dagger}F_{s}^{\dagger}  &  =F_{\uparrow+}^{\dagger}F_{\downarrow
+}~\text{,}\label{eq:kleinbil1}\\
F_{s\!f}F_{s}^{\dagger}  &  =F_{\uparrow-}^{\dagger}F_{\downarrow-}%
~\text{,}\label{eq:kleinbil2}\\
F_{s\!f}^{\dagger}F_{f}^{\dagger}  &  =F_{\uparrow+}^{\dagger}F_{\uparrow
-}~\text{,}\\
F_{c}^{\dagger}F_{s}^{\dagger}  &  =F_{\uparrow+}^{\dagger}F_{\uparrow
-}^{\dagger}~\text{.}%
\end{align}
These relations are consistent with the physical constraints given by the
variations in the fermionic numbers on the different sectors the Klein factors
act upon, and the choice of signs serves to fix otherwise arbitrary phases.
All other relations and their respective phases are now automatically fixed,
for instance,
\begin{align}
F_{s}^{\dagger}F_{f}^{\dagger}  &  =F_{\uparrow+}^{\dagger}F_{\downarrow
-}~\text{,}\\
F_{s}F_{c}^{\dagger}  &  =F_{\downarrow+}^{\dagger}F_{\downarrow-}^{\dagger
}~\text{,}\\
F_{f}^{\dagger}F_{s\!f}  &  =F_{\downarrow+}^{\dagger}F_{\downarrow-}~\text{,}
\label{eq:kleinbil7}%
\end{align}
can be deduced using the first four relations.

We want now to explore further the relations between old and new Klein
factors. For example, using Eqs.~(\ref{eq:kleinbil1}) and (\ref{eq:kleinbil2}%
), one can derive the following expressions:%
\end{subequations}
\begin{subequations}
\begin{align}
0=  &  \left\{  F_{\uparrow+},F_{\uparrow+}^{\dagger}F_{\downarrow+}\right\}
=\left\{  F_{\uparrow+},F_{s\!f}^{\dagger}F_{s}^{\dagger}\right\} \nonumber\\
&  =\left[  F_{\uparrow+},F_{s\!f}^{\dagger}\right]  F_{s}^{\dagger}%
+F_{s\!f}^{\dagger}\left\{  F_{\uparrow+},F_{s}^{\dagger}\right\}  ~\text{,}\\
0=  &  \left[  F_{\uparrow+},F_{\uparrow-}^{\dagger}F_{\downarrow-}\right]
=\left[  F_{\uparrow+},F_{s\!f}F_{s}^{\dagger}\right] \nonumber\\
&  =\left\{  F_{\uparrow+},F_{s\!f}\right\}  F_{s}^{\dagger}-F_{s\!f}\left\{
F_{\uparrow+},F_{s}^{\dagger}\right\}  ~\text{.}%
\end{align}
Premultiplying these equations by $F_{s\!f}$ and
$F_{s\!f}^{\dagger}$, respectively, and adding them up, one finds
\end{subequations}
\begin{equation}
F_{s\!f}F_{\uparrow+}F_{s\!f}^{\dagger}+F_{s\!f}^{\dagger}F_{\uparrow
+}F_{s\!f}=0~\text{.}%
\end{equation}
Using the different relations in Eqs.~(\ref{eq:kleinbil1})--(\ref{eq:kleinbil7}%
), one can generalize this relation to a generic relation between old ($o$)
and new ($n$) Klein factors:
\begin{equation}
F_{n}F_{o}F_{n}^{\dagger}+F_{n}^{\dagger}F_{o}F_{n}=0~\text{.}%
\end{equation}
Interestingly, such a relation is not consistent neither with
commutation nor with anticommutation relations between the old and new
Klein factors. This is not unexpected, since the two sets exist in
different Hilbert spaces and the physical identification between the
two is only through the relations among bilinears. While in
Kondo-type models this poses no problems because only Klein factor
bilinears enter in the Hamiltonians, this is not the case in
Anderson-type models,\cite{kotliar1996} for which we need to go beyond
bilinears.  Therefore, if we insist that the old and new Klein factors
should obey a relation of the type
\begin{equation}
F_{n}F_{o}=\alpha F_{o}F_{n}~\text{,}%
\end{equation}
we find that $\alpha^{2}=-1$ must hold. Thus, only semionic
commutation relations are consistent between old and new Klein
factors and there is still an arbitrariness in the phase that
needs to be fixed. Let us postulate
\begin{equation}
F_{\uparrow+}F_{c}=iF_{c}F_{\uparrow_{+}}~\text{;}%
\end{equation}
all other commutation relations can be deduced from this one and summarized
as
\begin{subequations}
\begin{align}
F_{\sigma\alpha}F_{\nu}  &  =i\theta_{\sigma\alpha}^{\nu}F_{\nu}%
F_{\sigma\alpha}~\text{,}\\
F_{\sigma\alpha}F_{\nu}^{\dagger}  &  =-i\theta_{\sigma\alpha}^{\nu}F_{\nu
}^{\dagger}F_{\sigma\alpha}~\text{,}%
\end{align}
where $\theta_{\sigma\alpha}^{c}=1$, $\theta_{\sigma\alpha}^{s}=\alpha
$, $\theta_{\sigma\alpha}^{f}=\sigma$, and $\theta_{\sigma\alpha}^{s\!f}%
=\sigma\alpha$.

A similar procedure can be used to verify that the new Klein factors and the
nondiagonal impurity Hubbard operators that enter in $H_{\text{hyb}}$ obey
\end{subequations}
\begin{equation}
F_{\nu}X_{\sigma\alpha}F_{\nu}^{\dagger}-F_{\nu}^{\dagger}X_{\sigma\alpha
}F_{\nu}=0~\text{,}%
\end{equation}
which is consistent only with standard commutation or anticommutation
relations between the operators and, on physical grounds, we take
them to anticommute.

Using the semionic commutation relations between the two sets of Klein
factors, it is easy to see that%
\begin{subequations}
\begin{align}
df  &  =ifd~\text{,}\\
f^{\dagger}d  &  =idf^{\dagger}~\text{.}%
\end{align}
Analogously, with respect to the spin-flavor band, we find
\end{subequations}
\begin{subequations}
\begin{align}
\psi_{s\!f}d  &  =-id\psi_{s\!f}~\text{,}\\
\psi_{s\!f}d^{\dagger}  &  =id^{\dagger}\psi_{s\!f}~\text{,}%
\end{align}
and
\end{subequations}
\begin{equation}
\left[  f,\psi_{s\!f}\right]  =\left[  f^{\dagger},\psi_{s\!f}\right]
=0~\text{.}%
\end{equation}
We say that $d$ is a \emph{relative semion} with respect to $f$ and
$\psi_{s\!f}$, while the latter two are \emph{relative bosons}.

On the other hand, the band fermions from the charge, spin and flavor sectors,
do not have simple commutation relations with the new impurity operators.
Those can be simplified by performing a unitary transformation of the Klein
factors belonging to that subspace. Namely, we define%
\begin{subequations}
\begin{align}
\tilde{F}_{s}  &  =e^{-i\frac{\pi}{2}X_{s}}F_{s}~\text{,}\\
\tilde{F}_{f}  &  =e^{-i\frac{\pi}{2}X_{f}}F_{f}~\text{,}\\
\tilde{F}_{c}  &  =e^{-i\frac{\pi}{4}X_{c}}F_{c}~\text{.}%
\end{align}
The redefined operators in the decoupled bands are now
\emph{relative fermions} with $d$, $f$, and among themselves. Notice also that
the Hamiltonian for the decoupled sectors is invariant under the redefinition.
What remains is to simplify further the commutation relations among the
spin-flavor operators and the impurity ones.

\subsubsection{Jordan-Wigner procedure}

The procedure we will use to simplify the commutation relations among
the impurity and spin-flavor operators is inspired on the
Jordan-Wigner treatment of spin chains. We have that $\psi_{s\!f}$,
$d$ and $f$ are all self-fermions while the pairs $(\psi_{s\!f},d)$
and $\left( d,f\right) $ are relative semions and, finally,
$(\psi_{s\!f},f)$ are relative bosons. For the sake of clarity, we
will split the Jordan-Wigner transformation into two parts. In the
first part, we shall change all operators into relative bosons, and in
the second part, we will perform a standard Jordan Wigner
transformation to turn them into relative fermions; that way the full
system will be a system of fermions.

Defining
\end{subequations}
\begin{subequations}
\begin{align}
\tilde{d}  &  =d~\text{,}\\
\tilde{f}  &  =e^{-i\pi n_{d}/2}f~\text{,}\\
\tilde{\psi}_{s\!f}  &  =e^{-i\pi n_{d}/2}\psi_{s\!f}~\text{,}%
\end{align}
(where $n_{d}=d^{\dagger}d$ and $n_{f}=f^{\dagger}f$), we see that all
operators are still self-fermions, but now their relative statistics is that
of relative bosons:
\end{subequations}
\begin{equation}
\lbrack\tilde{d},\tilde{f}]=[\tilde{d},\tilde{\psi}_{s\!f}]=[\tilde{f}%
,\tilde{\psi}_{s\!f}]=0~\text{.}%
\end{equation}
A set of operators that are relative bosons but all self-fermions, is
what is usually called core bosons. They are equivalent to spin-1/2
spins and the usual Jordan-Wigner treatment is available to turn them
into fermions. The second transformation is thus written as
\begin{subequations}
\begin{align}
f  &  =\tilde{f}~\text{,}\\
d  &  =e^{-i\pi n_{f}}\tilde{d}~\text{,}\\
\psi_{s\!f}  &  =e^{-i\pi(n_{f}+n_{d})}\tilde{\psi}_{s\!f}~\text{,}%
\end{align}
(notice that we have redefined $\psi_{s\!f}$, $d$ and $f$, and
that $n_{f}=n_{\tilde{f}}$ and $n_{d}=n_{\tilde{d}}$). While the
first Jordan-Wigner string was anchored at the $d$ site, the
second one is anchored at the $f$ site; this is why it is
physically more transparent to split the transformation into two
steps.

Finally, since the different fermionic species can be ordered such
that $\tilde{H}_{\text{hyb}}$ contains only nearest-neighbor hopping
terms, it can be seen that the Jordan-Wigner strings will not appear
explicitly in the Hamiltonian. After the double replacement, we find
\end{subequations}
\begin{multline}
\tilde{H}_{\text{hyb}}=V\left(  \psi_{s\!f}^{\dagger}d+d^{\dagger}\psi
_{s\!f}\right) \\
+\frac{V}{\sqrt{2\pi a}}\left[  \left(  f^{\dagger}-f\right)  \left(
d^{\dagger}+d\right)  \right]  ~\text{,}%
\end{multline}
where the operators are now the new ones and we were left with a system where
all operators are standard fermions (including those in the decoupled bands).

\subsection{Biresonant-level model}

\label{sec:bi-resonant}

\begin{figure}[ptb]
\begin{center}
\includegraphics[width=\columnwidth]{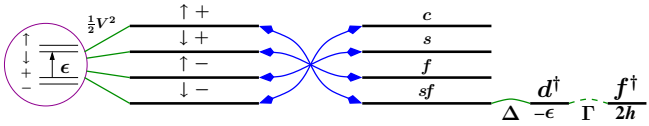}
\end{center}
\caption{Schematic representation of the mapping procedure between the
anisotropic two-channel Anderson model and the biresonant-level model.
The external field enters as a site energy for the $f^{\dagger}$ fermion and
has the effect of coupling its otherwise disconnected Majorana component.}%
\label{fig:bires}%
\end{figure}

Rewriting the remaining terms of the Hamiltonian in the language of
the new fermionic operators and choosing the anisotropy that would
make $\lambda_{\nu }=0\quad\forall\nu$, we find that the
transformations outlined above, in the limit of zero fields, give a
mapping to the following Fermi-Majorana biresonant-level model (up to
an additive constant, see also Fig.~\ref{fig:bires}):%
\noindent%
\begin{multline}
H_{\text{biRes}}=H_{0}^{s\!f}-\varepsilon~d^{\dagger}\!d
+\sqrt{2\Delta}\left[  \psi_{s\!f}^{\dagger}\left(  0\right)  d+d^{\dagger
}\psi_{s\!f}\left(  0\right)  \right]  +\\
+\sqrt{2\Gamma}\left(  f^{\dagger}-f\right)  \left(  d^{\dagger}+d\right)
~\text{,}%
\end{multline}
where $\varepsilon=\varepsilon_{s}-\varepsilon_{f}$,
$\Delta=V^{2}/2$ and $\Gamma=\Delta/2\pi a$. This is a purely
quadratic model, on which reintroducing the terms with non-zero
$\lambda_{\nu}$ would parametrize the
deviations from the solvable manifold (cf.~Refs.~%
%TCIMACRO{\TeXButton{clarke1993,sengupta1994}{\onlinecite
%{clarke1993,sengupta1994}}}%
%BeginExpansion
\onlinecite{clarke1993,sengupta1994}%
%EndExpansion
).

Rewriting the $H_{\text{field}}$ term after the mapping, we find
$H_{\text{field}}=H_{\text{field}}^{2}+H_{\text{field}}^{4}$ with%
%TCIMACRO{\TeXButton{%
%\begin{subequations}%
%}{\begin{subequations}}}%
%BeginExpansion
\begin{subequations}%
%EndExpansion%
\begin{align}
H_{\text{field}}^{2} & =h\left( 2n_{f}-1\right) -\lambda_{h}\left(
n_{d}+n_{f}-\tfrac{1}{2}\right) ~\text{,}\\ H_{\text{field}}^{4} &
=2\lambda_{h}n_{f}n_{d}~\text{,}%
\end{align}%
%TCIMACRO{\TeXButton{%
%\end{subequations}%
%}{\end{subequations}}}%
%BeginExpansion
\end{subequations}%
%EndExpansion
where we defined $\lambda_{h}=h_{f}-h_{s}$ and $2h=h_{f}+h_{s}$. Taking
$\lambda_{h}=0$ we find that the model is still quadratic even for non-zero
$h$. Nevertheless, we will see below that the presence of this term changes
the fixed point; we will talk in this case of a double-Fermi biresonant-level
model.

It is interesting to compare these two resonant-level models
between them and with the models that correspond to the solvable
points of the single- and two-channel Kondo models. In the latter
cases, one obtains a single-Majorana resonant-level model and a
single-Fermi resonant-level model, respectively. In contrast to
the Kondo models, the Anderson case requires two fermionic degrees
of freedom to parametrize the state of the impurity, which is now
more complex, since it allows for fluctuating states of valence. For
both the two-channel Kondo and Anderson models, in the absence of
external fields, ``half of a fermion'' coming from the impurity
decouples from the band and the models are naturally written in
terms of the Majorana components of the impurity degrees of
freedom. The Majorana component that exists completely decoupled
from the rest of the system is responsible for the
\textit{fractional} residual impurity entropy that we will discuss
below and constitutes a signature of the non-Fermi-liquid
properties of the fixed point. In both cases, the addition of an
external field has the effect of coupling the disconnected
Majorana component and driving the system toward a Fermi-liquid
type of fixed point, like the one of the single-channel Kondo
model.

\section{Stability of the Fixed Points}

\label{sec:stability}

As mentioned above, taking $\lambda_{x}=0$ (for $x=c$, $s$, $f$,
$s\!f$, $h$) defines a solvable manifold that contains a set of
fixed points. The important question to be addressed is how
generic and stable are those fixed points, how are they
parametrized and whether RG-flow trajectories starting outside the
manifold will flow to the same fixed points. The last
consideration is particularly important, since it addresses the
genericalness of the fixed points and the feasibility of
perturbative calculations in $\lambda_{x}$.

\subsection{Inside the soluble manifold}

In order to address these questions, we start by computing the
different Green functions. For that, we first write down the local
action for the impurity, in which the extended degrees of freedom from
the band were integrated out exactly; this procedure is justified in
order to study the interaction of the band with a local impurity that
exists only at, say, $x_{\text{imp}}=0$. In Nambu notation, we write
all the sums over positive Matsubara frequencies ($\omega_{n}$) only
and use the following spinor basis:
%TCIMACRO{\TeXButton{%
%\begin{widetext}%
%}{\begin{widetext}}}%
%BeginExpansion
\begin{widetext}%
%EndExpansion%
\begin{equation}
\Psi(\omega_{n})\equiv\left(
\begin{array}
[c]{cccccc}%
\psi_{s\!f}\left(  \omega_{n}\right)  & \psi_{s\!f}^{\dagger}\left(
-\omega_{n}\right)  & d\left(  \omega_{n}\right)  & d^{\dagger}\left(
-\omega_{n}\right)  & f\left(  \omega_{n}\right)  & f^{\dagger}\left(
-\omega_{n}\right)
\end{array}
\right)  ^{T}~\text{.}%
\end{equation}
The explicit form of the local action in the solvable manifold is
\begin{equation}
S=\frac{1}{\beta}\sum_{n\geq0}\Psi^{\dagger}(\omega_{n})A(\omega_{n}%
)\Psi(\omega_{n})~\text{,}%
\end{equation}
where $\beta=1/k_{\text{B}}T$ is the inverse temperature and
\begin{equation}
A(\omega_{n})=\left(
\begin{array}
[c]{cccccc}%
-2iv_{\text{F}} & 0 & \sqrt{2\Delta} & 0 & 0 & 0\\
0 & -2iv_{\text{F}} & 0 & -\sqrt{2\Delta} & 0 & 0\\
\sqrt{2\Delta} & 0 & -i\omega_{n}-\varepsilon & 0 & \sqrt{2\Gamma} &
-\sqrt{2\Gamma}\\
0 & -\sqrt{2\Delta} & 0 & -i\omega_{n}+\varepsilon & \sqrt{2\Gamma} &
-\sqrt{2\Gamma}\\
0 & 0 & \sqrt{2\Gamma} & \sqrt{2\Gamma} & -i\omega_{n}+2h & 0\\
0 & 0 & -\sqrt{2\Gamma} & -\sqrt{2\Gamma} & 0 & -i\omega_{n}-2h
\end{array}
\right)  ~\text{.}%
\end{equation}
The two-point Green functions are defined as $\beta
G(\omega_{n})\equiv \left\langle
\Psi(\omega_{n})\Psi^{\dagger}(\omega_{n})\right\rangle $ with
$G(\omega_{n})=A^{-1}(\omega_{n})$. Given the structure of the model,
it is convenient to work in terms of Majorana
fermions.\cite{maldacena1997} For that, we rotate each spinor
according to
\begin{equation}
\Psi_{M}(\omega_{n})\equiv R\Psi(\omega_{n})=\left(
\begin{array}
[c]{cccccc}%
\psi_{s\!f}^{\prime\prime}\left(  \omega_{n}\right)  & \psi_{s\!f}^{\prime
}\left(  \omega_{n}\right)  & d^{\prime\prime}\left(  \omega_{n}\right)  &
d^{\prime}\left(  \omega_{n}\right)  & f^{\prime\prime}\left(  \omega
_{n}\right)  & f^{\prime}\left(  \omega_{n}\right)
\end{array}
\right)  ^{T}~\text{,}%
\end{equation}%
%TCIMACRO{\TeXButton{%
%\end{widetext}%
%}{\end{widetext}}}%
%BeginExpansion
\end{widetext}%
%EndExpansion

\noindent where $R$ is a block diagonal matrix in which each block, given by%
\begin{equation}
r=\frac{1}{\sqrt{2}}%
\begin{pmatrix}
-i & i\\
1 & 1
\end{pmatrix}
~\text{,}%
\end{equation}
rotates a given Nambu doublet. With this convention, the Majorana Green
function matrix is $G_{M}(\omega_{n})=RG(\omega_{n})R^{\dagger}$ and its
expression in imaginary time is
\begin{align}
G_{M}(\tau-\tau^{\prime})  &  \equiv\left\langle \Psi_{M}(\tau)\Psi
_{M}^{\dagger}(\tau^{\prime})\right\rangle \nonumber\\
&  =\frac{1}{\beta}\sum_{n\geqslant0}G_{M}(\omega_{n})e^{-i\omega_{n}\left(
\tau-\tau^{\prime}\right)  }\nonumber\\
&  \qquad-G_{M}^{T}(\omega_{n})e^{i\omega_{n}\left(  \tau-\tau^{\prime
}\right)  }~\text{.}%
\end{align}

Let us now discuss the situation when $h=0$ (the case of $h\ne0$ will
be discussed in the Appendix). A direct calculation gives the
following results for the long-$\tau$ behavior of the different
diagonal Green functions:
\begin{subequations}
\begin{align}
\left\langle \psi_{s\!f}^{\prime\prime}(\tau)\psi_{s\!f}^{\prime\prime
}(0)\right\rangle  &  \sim\frac{1}{\tau}~\text{,} & \left\langle \psi
_{s\!f}^{\prime}(\tau)\psi_{s\!f}^{\prime}(0)\right\rangle  &  \sim\frac
{1}{\tau^{3}}~\text{,}\\
\left\langle d^{\prime\prime}(\tau)d^{\prime\prime}(0)\right\rangle  &
\sim\frac{1}{\tau}~\text{,} & \left\langle d^{\prime}(\tau)d^{\prime
}(0)\right\rangle  &  \sim\frac{1}{\tau^{3}}~\text{,}\\
\left\langle f^{\prime\prime}(\tau)f^{\prime\prime}(0)\right\rangle  &
\sim\frac{1}{\tau}~\text{,} & \left\langle f^{\prime}(\tau)f^{\prime
}(0)\right\rangle  &  \sim\operatorname*{sign}\tau~\text{.}%
\end{align}
Notice that different Majorana components of a given fermion exhibit
different asymptotics; this fact and the $1/\tau^{3}$ behavior of
certain Green functions were already stressed previously in the context
of the two-channel Kondo model.\cite{ye1997,ye1998} From these
correlators, one could read the following naive scaling dimensions (see
also Fig.~\ref{fig:bires_dims}): $[\psi_{s\!f}^{\prime\prime}%
]=[d^{\prime\prime}]=[f^{\prime\prime}]=1/2$, $[\psi_{s\!f}^{\prime
}]=[d^{\prime}]=3/2$, and $[f^{\prime}]=0$. The long time behavior of
all but one of the mixed correlation functions can be obtained using
these scaling dimensions; the exception being $\left\langle
\psi_{s\!f}^{\prime\prime}(\tau)d^{\prime\prime}(0)\right\rangle
\sim1/\tau^{3}$.

\begin{figure}[ptb]
\begin{center}
\includegraphics[width=.8\columnwidth]{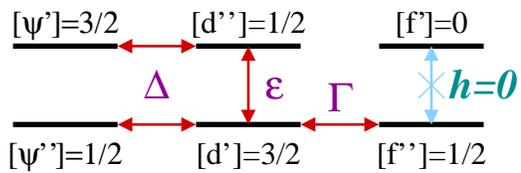}
\end{center}
\caption{Connectivity graph, in terms of Majorana operators, for the
biresonant-level model. The scaling dimensions of the fields are seen
to alternate between $1/2$ and $3/2$, with the exception of $f'$ that
stays decoupled (for $h=0$) and has zero scaling dimension.}%
\label{fig:bires_dims}%
\end{figure}

The scaling dimensions above are not compatible with a standard
renormalization group scheme that proceeds by integrating energy
shells while keeping certain coefficients of the quadratic action
constant, cf.~Ref.~%
%TCIMACRO{\TeXButton{shankar1994}{\onlinecite{shankar1994}}}%
%BeginExpansion
\onlinecite{shankar1994}%
%EndExpansion
. Instead, we follow the ideas of Ref.~%
%TCIMACRO{\TeXButton{moustakas1996}{\onlinecite{moustakas1996}} }%
%BeginExpansion
\onlinecite{moustakas1996}
%EndExpansion
and introduce the Majorana Fermi velocities ($v_{\psi^{\prime}}$,
$v_{\psi^{\prime\prime}}$) as well as constants multiplying the frequency in
the Berry terms for the impurity ($g_{d^{\prime}}$, $g_{d^{\prime\prime}}$,
$g_{f^{\prime}}$, $g_{f^{\prime\prime}}$) and let all of them vary.
Specifically, we proceed along the standard steps\cite{shankar1994} but
insisting that the fields should scale with the dimensions we determined from
the Green functions, this yields the following RG-flow equations for the
different couplings:%
\end{subequations}
\begin{subequations}
\begin{align}
\frac{dg_{d^{\prime}}}{d\ell}  &  =-3g_{d^{\prime}}~\text{,} & \frac
{dg_{d^{\prime\prime}}}{d\ell}  &  =-g_{d^{\prime\prime}}~\text{,}\\
\frac{dg_{f^{\prime}}}{d\ell}  &  =0~\text{,} & \frac{dg_{f^{\prime\prime}}%
}{d\ell}  &  =-g_{f^{\prime\prime}}~\text{,}\\
\frac{dv_{\psi^{\prime}}}{d\ell}  &  =-2v_{\psi^{\prime}}~\text{,} &
\frac{dv_{\psi^{\prime\prime}}}{d\ell}  &  =0~\text{,}\\
\frac{d\Delta}{d\ell}  &  =-\Delta~\text{,} & \frac{d\varepsilon}{d\ell}  &
=-\varepsilon~\text{,}\\
\frac{d\Gamma}{d\ell}  &  =-\Gamma~\text{,} & \frac{dh}{d\ell}  &  =\frac
{1}{2}h~\text{.}%
\end{align}
We find that most couplings are RG-irrelevant but for $v_{\psi^{\prime\prime}%
}$ and $g_{f^{\prime}}$ that are RG-marginal. In here, we are considering
$h=0$, but we find that if a field were to be generated it would be
RG-relevant. It is important to remark that the the quotient $\varepsilon
/\Delta$ obeys%
\end{subequations}
\begin{equation}
\frac{d}{d\ell}\left(  \frac{\varepsilon}{\Delta}\right)  =0
\end{equation}
and is also marginal. This agrees with the results of Bethe
ansatz\cite{bolech2002,bolech2005a} and boundary conformal field
theory\cite{johannesson2003} that find a line of non-Fermi-liquid fixed points
that, in the microscopic theory, are parametrized by the $\varepsilon/\Delta$
ratio and interpolate between spin and flavor two-channel Kondo behaviors.

\subsection{Outside the soluble manifold}

Collecting the terms in the refermionized Hamiltonian that takes us away from
the soluble manifold, we have%
\begin{align}
H_{\lambda}  &  =-\lambda_{c}\rho_{c}-\lambda_{s\!f}\rho_{s\!f}-\lambda
_{s}\rho_{s}\nonumber\\
&  +2\lambda_{c}\rho_{c}n_{d}+2\lambda_{s\!f}\rho_{s\!f}n_{d}+\lambda_{f}%
\rho_{f}n_{d}\nonumber\\
&  +\lambda_{s}\rho_{s}n_{d}+2\lambda_{s}\rho_{s}n_{f}\nonumber\\
&  -2\lambda_{f}\rho_{f}n_{d}n_{f}-2\lambda_{s}\rho_{s}n_{d}n_{f}\nonumber\\
&  -\lambda_{h}\left(  n_{d}+n_{f}-\tfrac{1}{2}\right)  +2\lambda_{h}%
n_{f}n_{d}~\text{.}%
\end{align}
In order to determine the stability of the $h=0$ fixed points, we
proceed to study the RG-relevance of the different terms in
$H_{\lambda}$. The first three terms are just chemical potential terms
for the conduction electrons. The charge and spin ones pertain only to
bands that are decoupled from the impurity and are marginal operators,
since the dimension of the respective fields is $1/2$, as dictated by
their Berry phases. On the other hand, the spin-flavor one is
irrelevant as can be seen by simple power counting.

For the operators in the second, third and fourth lines, we get the
following dimensions by summing the dimensions of their constituents
$[\rho_{c}n_{d}]=[\rho_{f}n_{d}]=[\rho_{s}n_{d}]=3$,
$[\rho_{s\!f}n_{d}]=4$, $[\rho_{s}n_{f}]=3/2$ and
$[\rho_{f}n_{d}n_{f}]=[\rho_{s}n_{d}n_{f}]=7/2$. Hence, all of them
are RG-irrelevant. None of these terms can generate a local field term
for the impurity. In particular, $\psi_{c}$, $\psi_{s}$ and $\psi_{f}$
are decoupled (and the average of the normal-ordered densities
$\rho_{c}$, $\rho_{s}$ and $\rho_{f}$ is zero). Eventually, the
operator $\rho_{s\!f}n_{d}$ may give a correction to $\varepsilon$,
thus taking the system along the line of fixed points.

An interesting situation appears when one starts from the fixed point
in the absence of a magnetic field ($h_s=0$) and considers the effect of a
small $\lambda_{h}\neq0$ (this is realized when one starts with
$\lambda_{h}=2h$, or in other words, with $h_f\ne0$). Then one must
consider carefully the effect of the last term in $H_{\lambda}$, with
dimension $[n_{f}n_{d}]=5/2$ and therefore, in principle,
RG-irrelevant.  This term is actually classified as dangerously
irrelevant, as it generates a potential field for $f$ (\textit{i.e.},
an RG-relevant $h$ term). It thus couples the remaining Majorana
component, $f^{\prime}$, and changes the fixed point.

Indeed, when adding a magnetic field the fixed point changes (see
Appendix). The chemical potentials for $\psi_{sf}$ and $d$ become
marginal, and the operators with four and six fermions are now all
irrelevant. The parameters $\varepsilon/\Delta$ and $h/\Gamma$
parametrize the new set of Fermi-liquid fixed points.

\section{Thermodynamics with Finite Field}

\label{sec:thermodynamics}

In order to extract the impurity thermodynamics when the model
parameters are on the solvable manifold, we can resort to an exact
calculation of the free energy. The impurity free energy can be
straightforwardly calculated by means of Pauli's trick of integration
over the coupling constants (an alternative way is to first derive
an effective action for the impurity\cite{gan1995}), namely,%
\begin{equation}
\Omega-\Omega_{0}=\int_{0}^{1}\frac{d\xi}{\xi}~\left\langle \xi\left(
H_{\text{biRes}}-H_{0}^{s\!f}\right)  \right\rangle _{\xi}~\text{.}%
\end{equation}
The $\xi$ subindex indicates that the mean values should be computed using an
action in which all couplings involving the impurity were multiplied by the
dimensionless parameter $\xi$, and $\Omega_{0}$ is the impurity free energy
when $\xi=0$. After computing the mean values using $G_{\xi}\left(  \omega
_{n}\right)  $, one arrives at the expression%
\begin{equation}
\Omega-\Omega_{0}=-\int_{0}^{1}d\xi~\frac{1}{\beta}\sum_{n\geq0}\frac
{\partial_{\xi}D\left(  \omega_{n},\xi\right)  }{D\left(  \omega_{n}%
,\xi\right)  }~\text{,}%
\end{equation}
where%
\begin{align}
D(\omega_{n},\xi)  &  \equiv-\frac{1}{4}\det A_{\xi}\left(  \omega_{n}\right)
=4h^{2}\Delta^{2}\xi^{6}\nonumber\\
&  +\left[  \Delta\omega_{n}\left(  8\Gamma+\Delta\omega_{n}\right)
+4h^{2}\left(  \varepsilon^{2}+2\Delta\omega_{n}\right)  \right]  \xi
^{4}\nonumber\\
&  +\omega_{n}^{2}\left(  4h^{2}+8\Gamma+\varepsilon^{2}+2\Delta\omega
_{n}\right)  \xi^{2}+\omega_{n}^{4}~\text{.}%
\end{align}
As a function of $\omega_{n}$, this is a fourth order polynomial
with roots that are parametric functions of $\xi$:
$\omega_{k}\left(  \xi\right)  $, $k=0,\ldots,3$. Using the
factorized form of the polynomial in terms of its roots and
introducing a suitable regularization, we arrive at the following
expression:
\begin{align}
\Omega-\Omega_{0}  &  =\int_{0}^{1}d\xi\frac{1}{\beta}\sum_{n\geq0}\sum
_{k}\frac{\partial_{\xi}\omega_{k}(\xi)}{\omega_{n}-\omega_{k}(\xi
)}\nonumber\\
&  =\int_{0}^{1}d\xi\frac{1}{\beta}\sum_{n\geq0}\sum_{k}\partial_{\xi}%
\omega_{k}(\xi)\int_{-\Lambda}^{\omega_{k}(\xi)}d\mu\frac{1}{\left[
\omega_{n}-\mu\right]  ^{2}}~\text{.}%
\end{align}
We verify that $\omega_{k}(0)=0$ and call $\omega_{k}\equiv\omega_{k}\left(
1\right)  $. Exchanging the sum in $n$ and the integral in $\mu$, we arrive at
the final expression:
\begin{multline}
\Omega-\Omega_{0}=\sum_{k}\frac{1}{2\pi}\left[  \psi\left(  \frac{1}{2}%
+\frac{\Lambda\beta}{2\pi}\right)  \omega_{k}\right. \\
+\left.  \frac{2\pi}{\beta}\ln\Gamma\left(  \frac{1}{2}-\frac{\omega_{k}\beta
}{2\pi}\right)  -\frac{2\pi}{\beta}\ln\Gamma\left(  \frac{1}{2}\right)
\right]  ~\text{,}%
\end{multline}
where $\psi\left(  z\right)  \equiv\partial_{z}\ln\Gamma\left(  z\right)  $ is
the digamma function. Note that the free energy needs the $\Lambda$ regulator,
but in all the derived thermodynamic quantities we will be able to take the
the limit $\Lambda\rightarrow+\infty$ and obtain finite results.%

%TCIMACRO{\TeXButton{fig:entropy}{\begin{figure*}
%\begin{center}
%\includegraphics[width=\textwidth]{entropy}
%\end{center}
%\caption
%{Impurity contribution to the entropy as a function of temperature for different values of the
%symmetric external field $h$. Temperature and field are both measured in units of $\Delta
%$. Different curves
%correspond to different values of $\varepsilon/\Delta
%$. The top left panel corresponds to zero field and
%the other ones to finite values of the field that increase hundred-fold between panels, from $h=10^{-5}%
%\Delta$
%until $h=10^{3}\Delta$.}
%\label{fig:entropy}
%\end{figure*}}}%
%BeginExpansion
\begin{figure*}
\begin{center}
\includegraphics[width=\textwidth]{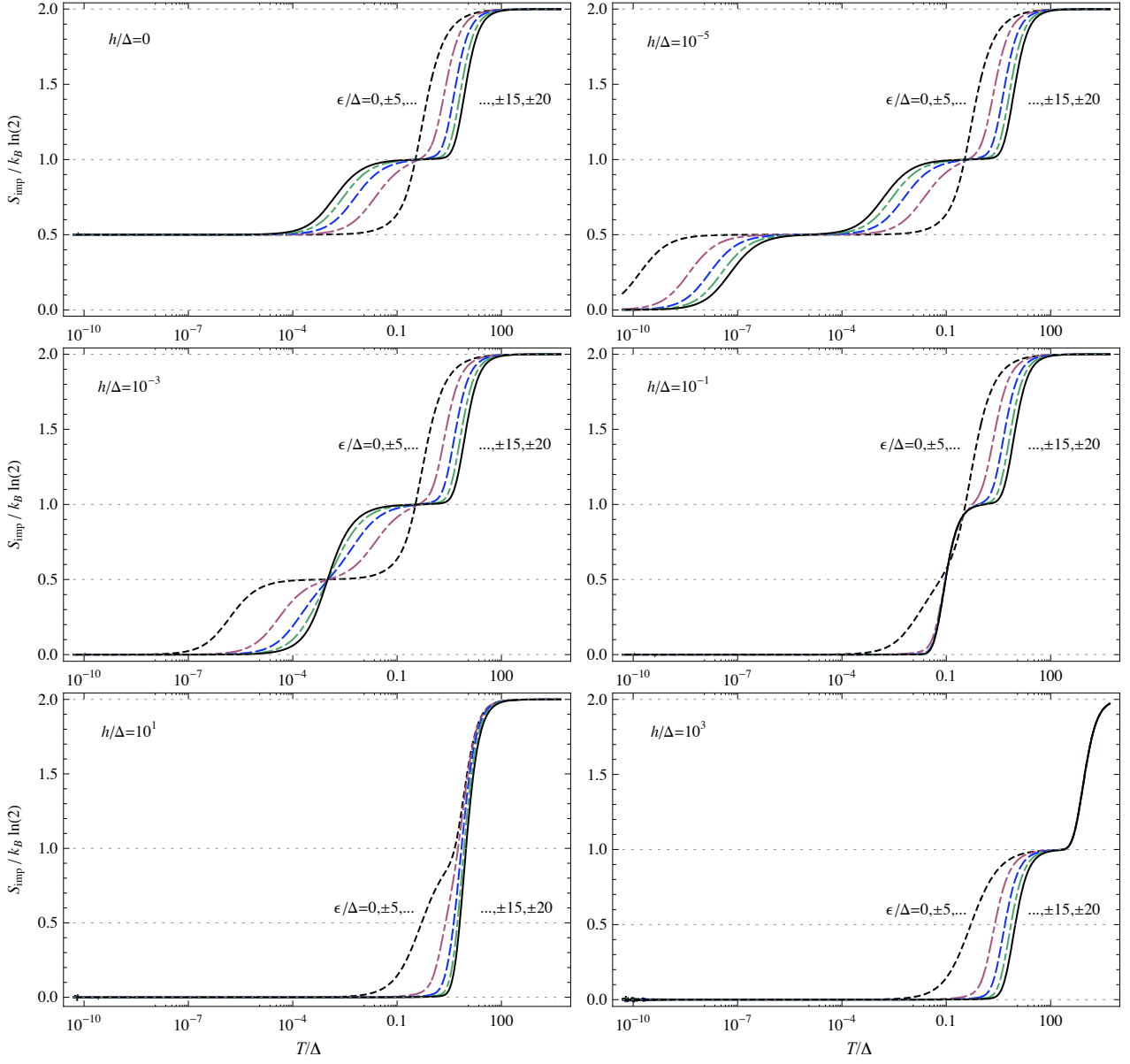}
\end{center}
\caption
{Impurity contribution to the entropy as a function of temperature for
different values of the symmetric external field $h$. Temperature and
field are both measured in units of $\Delta $. Different curves
correspond to different values of $\varepsilon/\Delta $ and
$\Gamma=\Delta/2\pi$. The top left panel corresponds to zero field and
the other ones to finite values of the field that increase
hundredfold between panels, from $h=10^{-5}%
\Delta$
until $h=10^{3}\Delta$.}
\label{fig:entropy}
\end{figure*}%
%EndExpansion

In particular, the impurity entropy is given by $S-S_{0}=\sum_{k}s\left(
z_{k}\right)  $, with $z_{k}=-\beta\omega_{k}/2\pi$ and%
\begin{equation}
s\left(  z\right)  =z\left[  \psi\left(  \frac{1}{2}+z\right)  -1\right]
-\ln\Gamma\left(  \frac{1}{2}+z\right)  +\frac{1}{2}\ln\pi~\text{.}%
\end{equation}
Remarkably, the function $s\left(  z\right)  $ is the same that is found for
the entropy of the two-channel Kondo model on the Emery-Kivelson line.
\cite{fabrizio1995} This is not completely unexpected, since Kondo is the
low-energy effective theory for most part of the parameter regime of the
two-channel Anderson model. In Fig.~\ref{fig:entropy}, we show the impurity
contribution to the entropy as a function of temperature for different values
of the local field $h$.

Let us first discuss the $h=0$ case shown on the top left panel of the
figure. In this case, one of the roots is zero ($\omega_{3}=0$) and
there is a residual entropy of
$S_{\text{imp}}=k_{\text{B}}\ln\sqrt{2}$ that signals a
non-Fermi-liquid set of fixed points. The finite temperature physics
is governed by the other three roots.\cite{bolech2006a} We find that,
within the range of parameters of physical relevance, one of the roots
($\omega_{0}$) is real while the other two ($\omega_{1,2}$) are
complex conjugate of each other.  This makes it natural to identify
the roots with the Kondo and Schottky energy scales in the following
way:%
\begin{align}
k_{\text{B}}T_{\text{K}}  &  =-\omega_{0}/2\pi~\text{,}\\
k_{\text{B}}T_{\text{S}}  &  =\left\vert \omega_{1}\right\vert /2\pi~\text{.}%
\end{align}
The Kondo scale corresponds to a jump in the entropy of hight $k_{\text{B}}%
\ln\sqrt{2}$, while the Schottky scale is associated with a jump
that is twice as large. For small $\left\vert
\varepsilon\right\vert $, the two quenching steps coincide;
whereas as $\left\vert \varepsilon\right\vert /\Delta$ grows, the
Kondo temperature decreases while the Schottky temperature
increases. Evidently, over the soluble manifold, the dependence of
all scales on the microscopic parameters is algebraic. This
contrasts with the isotropic case, for which the Kondo scale has
exponential dependence on $\left\vert \varepsilon\right\vert
/\Delta$.\cite{bolech2005a} This behavior is seen also in the
single- and two-channel Kondo models, for which it is found that
the exponential dependence of the Kondo scale is a property of the
isotropic models only and the functional dependence crosses over
to algebraic in the Toulouse and Emery-Kivelson limits,
respectively.\cite{tsvelik1983,zarand2002} Another characteristic
of the solvable limit is that the $\ln T$ dependencies in the
impurity susceptibilities or in the specific heat coefficient are
changed into power laws. It is known that the logarithms can be
recovered
using perturbation theory in $\lambda_{\nu}$ (cf.~Refs.~%
%TCIMACRO{\TeXButton{emery1992,clarke1993,sengupta1994}{\onlinecite
%{emery1992,clarke1993,sengupta1994}}}%
%BeginExpansion
\onlinecite{emery1992,clarke1993,sengupta1994}%
%EndExpansion
), as it reintroduces the leading RG-irrelevant operators that are
absent from the soluble manifold. For the leading contribution, the
three operators with scaling dimension $3$ (\textit{i.e.},
$\rho_cn_d$, $\rho_sn_d$, and $\rho_fn_d$) can be included using
second-order perturbation theory in the couplings. All three of them
mix one of the decoupled bands and the impurity in a way that is the
direct generalization of what happens in the case of the two-channel
Kondo model.\cite{sengupta1994} Such a procedure should recover, for
instance, not only the results for the specific heat and
susceptibilities but also the Wilson ratio that was discussed
already in Ref.~\onlinecite{johannesson2003}.

When $h\neq0$, $\omega_{3}$ becomes non-zero as well and grows with
$h$. This introduces a third quenching step (as it can be seen, for
instance, in the top right panel of the figure) in which the entropy
goes from $k_{\text{B}}\ln\sqrt{2}$ to zero, signaling the system
flowing away from the non-Fermi-liquid fixed point. This is in
accordance with the result of the previous section, in which we found
$h$ to be a relevant perturbation. This process defines a third energy
scale,%
\begin{equation}
k_{\text{B}}T_{h}=-\omega_{3}/2\pi~\text{,}%
\end{equation}
that grows with $h$ until reaching the same value as
$T_{\text{K}}$, --which happens at different values of the field
for different values of $\left\vert \varepsilon\right\vert$
(taking place first for larger values, see the middle left panel
in Fig.~\ref{fig:entropy}). As these two energy scales merge, the
respective roots become a complex-conjugate pair, a single
quenching step of the same height as the Schottky one emerges and
the transition width (\textit{i.e.}, the width of the
corresponding anomaly in the specific heat) narrows. As $h$
increases further, the three scales first become degenerate and
then cross each other. In the bottom right panel, the two scales
have already crossed, the higher temperature step is given by
$T_{\text{K}}=T_{h}$ and is insensitive to the value of
$\left\vert \varepsilon\right\vert /\Delta $, while the lower one
continues to correspond to the Schottky transition and shows
the same type of dependence in $\left\vert \varepsilon\right\vert
/\Delta$ as for lower fields. These results share many properties
with those for the isotropic case, but they are specific to the
case of a symmetric filed ($\lambda_{h}=0$) and exhibit important
differences with the case when only $h_{s}$ or $h_{f}$ is applied
to the system (cf.~Fig.~7 in Ref.~%
%TCIMACRO{\TeXButton{bolech2005a}{\onlinecite{bolech2005a}}}%
%BeginExpansion
\onlinecite{bolech2005a}%
%EndExpansion
).%

%TCIMACRO{\TeXButton{fig:valence}{\begin{figure}
%\begin{center}
%\includegraphics[width=\columnwidth]{valence}
%\end{center}
%\caption
%{Impurity charge valence as a function of temperature. The curves are displayed for $h=0$,
%the only effect of finite fields is to wash out the overshoot that happens for small values of $¦\varepsilon
%¦/\Delta$
%and to shift the zero-temperature values of $n_c$(see text).}
%\label{fig:valence}
%\end{figure}}}%
%BeginExpansion
\begin{figure}
\begin{center}
\includegraphics[width=\columnwidth]{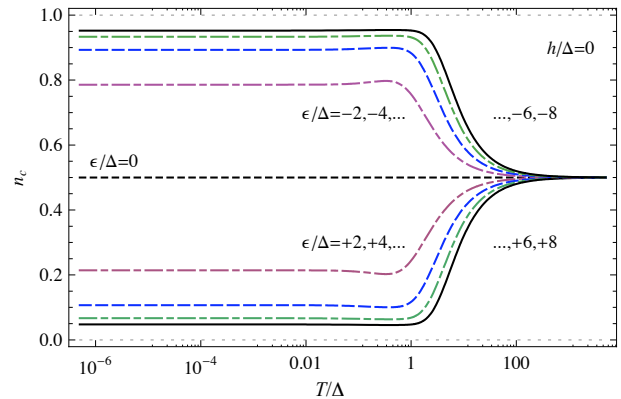}
\end{center}
\caption
{Impurity charge valence as a function of temperature. The curves are
displayed for $h=0$, the only effect of finite fields is to wash out
the overshoot that happens for small values of $¦\varepsilon ¦/\Delta$
and to shift (away from $1/2$) the zero-temperature values of
$n_c$ (see text).}
\label{fig:valence}
\end{figure}%
%EndExpansion

\begin{figure}
\begin{center}
\includegraphics[width=\columnwidth]{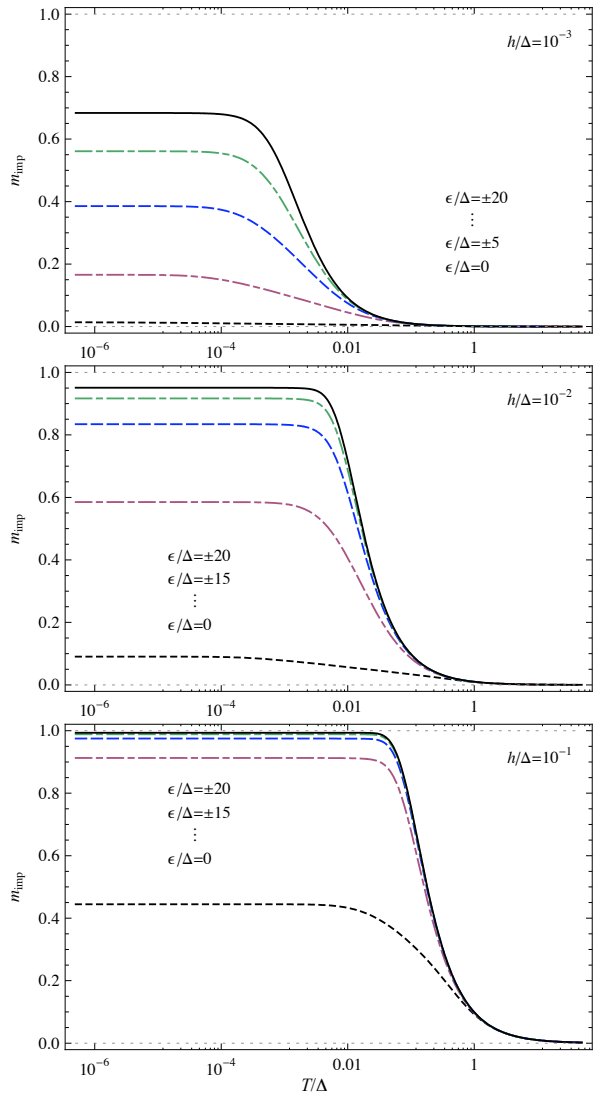}
\end{center}
\caption
{Impurity magnetization versus temperature. From top to bottom,
the three panels show three different
values of $h/\Delta=10^{-3}, 10^{-2}, 10^{-1}%
$. Curves for different values of $\varepsilon/\Delta$ are shown
and the values are indicated in the figure.}
\label{fig:magnetization}
\end{figure}

Other thermodynamic quantities of interest are the impurity charge valence
$n_{c}=\partial\Omega/\partial\varepsilon$ and the magnetization
$m_{\text{imp}}=-\partial\Omega/\partial h$. They are explicitly given by the
expressions:%
\begin{subequations}
\begin{align}
n_{c}-n_{c,0}  &  =\frac{1}{\beta}\sum_{k}\psi\left(  \frac{1}{2}%
+z_{k}\right)  \partial_{\varepsilon}z_{k}~\text{,}\\
m_{\text{imp}}-m_{\text{imp},0}  &  =-\frac{1}{\beta}\sum_{k}\psi\left(
\frac{1}{2}+z_{k}\right)  \partial_{h}z_{k}~\text{,}%
\end{align}
\end{subequations} where the derivatives of the roots can be
expressed in closed form using the
identity%
\begin{equation}
\partial_{\varepsilon,h}\omega_{k}=-\frac{\left.  \partial_{\varepsilon
,h}D(\omega,1)\right\vert _{\omega=\omega_{k}}}{\prod_{j\neq k}\left(
\omega_{k}-\omega_{j}\right)  }~\text{.}%
\end{equation}
Plots of these two quantities are displayed in
Figs.~\ref{fig:valence} and \ref{fig:magnetization}, respectively.

The variability of the impurity valence is an aspect inherent to Anderson-type
models and of relevance, for instance, in the context of quantum dots and
other mesoscopic systems that might allow direct measurements of the valence
states via changes in capacitance.\cite{bolech2005b} The valence starts as
$n_{c,0}=1/2$ at high temperature ($T\gg T_{\text{S}}$) independently of the
values of $\varepsilon$ and $h$, and evolves as the temperature is lowered to
attain a certain zero-temperature value $n_{c}^{0}\equiv n_{c}\left(
\varepsilon,h\right)  _{T=0}$ that characterizes the particular fixed point
that the system reaches. The quenching of the valence fluctuations coincides
with the first quenching of the entropy and takes place at the characteristic
scale $T_{\text{S}}$. Subtle aspects of the small $\left\vert \varepsilon
\right\vert $ curves, such as the ``overshoot'' of the curves at intermediate
temperatures, $T\lesssim T_{\text{S}}$, are generic and are also
present in the exact solution of the isotropic
model.\cite{bolech2005a,bolech2006a,bolech2006b} This feature exists
for $h$ zero or small and disappears as $h$ becomes of the order of
$\Delta$. Comparisons with the isotropic model show that, despite the
differences in the infrared physics, the isotropic and the anisotropic
models share the same generic ultraviolet physics.\cite{bolech2006a}
This is to be expected, since the physics at higher energies tends
to be dominated by local fluctuations.%

%TCIMACRO{\TeXButton{fig:magnetization}{\begin{figure}
%\begin{center}
%\includegraphics[width=\columnwidth]{magnetization}
%\end{center}
%\caption
%{Impurity magnetization versus temperature. From top to bottom the three panels show three different
%values of $h/\Delta=10^{-3}, 10^{-2}, 10^{-1}%
%$. Curves for different values of $\varepsilon/\Delta$ are shown
%and the value are indicated in the figure.}
%\label{fig:magnetization}
%\end{figure}}}%
%BeginExpansion
%EndExpansion

The curves of finite-field impurity magnetization as a function of temperature
are in some ways similar to those of the charge valence. At high temperatures,
the magnetization is zero, and it acquires a field-dependent finite value as
the temperature becomes $T\lesssim T_{h}$ (the magnetization does not reach
the maximum value of $1$ due to the hybridization of the impurity with the
conduction band). In this case, curves for positive and negative values of
$\varepsilon$ are degenerate. The quantities $m_{\text{imp}}^{0}\equiv
m_{\text{imp}}\left(  \varepsilon,h\right)  _{T=0}$ and $n_{c}^{0}$ label the
set of finite-field fixed points of the model. As we mentioned above, these
fixed points are of the local Fermi-liquid type.

\section{Conclusion}

\label{sec:conclusion}

We have shown how the addition of exchange terms that provide anisotropy that
breaks the $SU\left(  2\right)$ symmetry of the two-channel Anderson model
in the spin and flavor sectors, by singling out an easy access for each of
them, allows to identify a manifold in coupling constant space over which the
model becomes quadratic and thus exactly solvable. The solvable theory is
obtained in the form of a mapping to an anusual type of resonant-level model
that includes two resonant levels and we thus call it a biresonant-level
model. Similar mappings exist for the Kondo model in the case of one and two
channels (and are known to be impossible for larger number of
channels\cite{tsvelik1995}). But a simple mapping of this type is new for a
model like the two-channel Anderson model that incorporates the physics of
valence fluctuations. Equivalent mappings for the single-channel Anderson
model are elusive (cf.~Ref.~%
%TCIMACRO{\TeXButton{kotliar1996}{\onlinecite{kotliar1996}}}%
%BeginExpansion
\onlinecite{kotliar1996}%
%EndExpansion
), which indicates that the case of the two-channel model
constitutes a singular example. Since the mapping captures, in
particular, the mixed valence regime of the model, --in which
electrons fluctuate from the impurity to the band and vice versa--,
it is of utmost importance to verify that such physics is
rigorously captured by including the Klein factors and showing how
they take part in the transformations that are required for the
mapping. A careful study of the Klein factors opens the door for
future calculations of correlation functions using the solvable
points as starting ground. The representativity of calculations
that start from the biresonant-level model requires that such
model flows into generic fixed points representative of the
physics of the original model. We performed an analysis of the
stability of deviations from the locus of solvable points in
parameter space and established that perturbation theory on such
deviations is valid. The advantage of starting from the mapped
model is evident: closed analytical expressions can be derived for
the full temperature crossovers of the different quantities of
interest; this sets the approach apart from other non-perturbative
techniques applied previously to the two-channel Anderson model.

To illustrate the versatility, power and simplicity of the use of
the biresonant-level model to do calculations, we presented
several results of thermodynamic quantities that involve the
impurity. Namely, we gave results for the impurity entropy,
valence and magnetization as functions of temperature for
different values of $\varepsilon$ and $h$. These quantities
illustrate the existence of several energy scales in the model
that signal the transitions between different regimes. It is
remarkable that all the crossovers can be calculated explicitly in
closed analytic expressions, something not possible with other
types of calculations.

For the followup, it would be interesting to exploit the precise
mapping (that includes all the details such as Klein factors,
decoupled bands, etc.) to compute dynamic correlation functions
involving the impurity. Another interesting direction would be the
study of the finite-size crossover spectrum of the model, which would
allow to establish instructive connections with different
renormalization group schemes. Finally, we are in a position to, for
instance, consider the behavior of more than one impurity
(cf.~Ref.~\onlinecite{ferrero2007}) and, in particular, the case of
one-dimensional two-channel Anderson lattices
(cf.~Ref.~\onlinecite{schauerte2005}). The latter would be interesting
in order to determine if lattice effects are able to reverse the sign
of the prefactor in the leading $\sqrt{T}$ term in the
resistivity,\cite{johannesson2005} which
would be required in order to match the experimental results for
thoriated \textrm{UBe}$_{13}$.\cite{aliev1995,dickey1997}

\begin{acknowledgments}
We acknowledge discussions with J. von Delft, T. Giamarchi, G. Zar\'and
and M. Zvonarev. This work was partly supported by the Swiss National
Science Foundation under MaNEP and Division II, the W.~M.~Keck Foundation
and DARPA.
\end{acknowledgments}

\appendix

\section*{Appendix: Study of the finite-field fixed points}

\label{sec:fieldRG}

The long-$\tau$ asymptotics change dramatically when a non-zero external field
is present ($h\neq0$). The $f^{\prime}$ Majorana component of $f$ is now
coupled to the rest of the system and we must recompute all the Green
functions. For the diagonal ones, we find%
\begin{subequations}
\begin{align}
\left\langle \psi_{s\!f}^{\prime\prime}(\tau)\psi_{s\!f}^{\prime\prime}(0)\right\rangle  &
\sim\frac{1}{\tau}~\text{,} & \left\langle \psi_{s\!f}^{\prime}(\tau)\psi_{s\!f}^{\prime
}(0)\right\rangle  &  \sim\frac{1}{\tau}~\text{,}\\
\left\langle d^{\prime\prime}(\tau)d^{\prime\prime}(0)\right\rangle  &
\sim\frac{1}{\tau}~\text{,} & \left\langle d^{\prime}(\tau)d^{\prime
}(0)\right\rangle  &  \sim\frac{1}{\tau}~\text{,}\\
\left\langle f^{\prime\prime}(\tau)f^{\prime\prime}(0)\right\rangle  &
\sim\frac{1}{\tau^{3}}~\text{,} & \left\langle f^{\prime}(\tau)f^{\prime
}(0)\right\rangle  &  \sim\frac{1}{\tau}~\text{,}%
\end{align}
from where we read the new naive scaling dimensions:
$[\psi_{s\!f}^{\prime\prime}%
]=[\psi_{s\!f}^{\prime}]=[d^{\prime\prime}]=[d^{\prime}]=[f^{\prime}]=1/2$
and $[f^{\prime\prime}]=3/2$. For the nondiagonal Green functions,
the following ones have long-$\tau$ behaviors that do not match the
expectations from the naive scalings:
\end{subequations}
\begin{subequations}
\begin{align}
\left\langle \psi_{s\!f}^{\prime}(\tau)\psi_{s\!f}^{\prime\prime}(0)\right\rangle  &
\sim\dfrac{1}{\tau^{2}}~\text{,} & \left\langle \psi_{s\!f}^{\prime}(\tau
)d^{\prime\prime}(0)\right\rangle  &  \sim\dfrac{1}{\tau^{2}}~\text{,}\\
\left\langle \psi_{s\!f}^{\prime\prime}(\tau)d^{\prime}(0)\right\rangle  &
\sim\dfrac{1}{\tau^{2}}~\text{,} & \left\langle \psi_{s\!f}^{\prime\prime}%
(\tau)f^{\prime\prime}(0)\right\rangle  &  \sim\dfrac{1}{\tau^{3}}~\text{,}\\
\left\langle \psi_{s\!f}^{\prime\prime}(\tau)f^{\prime}(0)\right\rangle  &
\sim\dfrac{1}{\tau^{2}}~\text{,} & \left\langle d^{\prime\prime}%
(\tau)d^{\prime}(0)\right\rangle  &  \sim\dfrac{1}{\tau^{2}}~\text{,}\\
\left\langle d^{\prime\prime}(\tau)f^{\prime\prime}(0)\right\rangle  &
\sim\dfrac{1}{\tau^{3}}~\text{,} & \left\langle d^{\prime\prime}%
(\tau)f^{\prime}(0)\right\rangle  &  \sim\dfrac{1}{\tau^{2}}~\text{.}%
\end{align}
All of them decay faster than what the naive use of the scaling dimensions of
the fields would indicate. Thus, in order to obtain the RG flow of the
nondiagonal couplings, one should, all the same, employ the naive dimensions.
The resulting flow equations are as follows:
\end{subequations}
\begin{subequations}
\begin{align}
\frac{dg_{d^{\prime}}}{d\ell}  &  =-g_{d^{\prime}}~\text{,} & \frac
{dg_{d^{\prime\prime}}}{d\ell}  &  =-g_{d^{\prime\prime}}~\text{,}\\
\frac{dg_{f^{\prime}}}{d\ell}  &  =-g_{f^{\prime}}~\text{,} & \frac
{dg_{f^{\prime\prime}}}{d\ell}  &  =-3g_{f^{\prime\prime}}~\text{,}\\
\frac{dv_{\psi^{\prime}}}{d\ell}  &  =0~\text{,} & \frac{dv_{\psi
^{\prime\prime}}}{d\ell}  &  =0~\text{,}\\
\frac{d\Delta}{d\ell}  &  =0~\text{,} & \frac{d\varepsilon}{d\ell}  &
=0~\text{,}\\
\frac{d\Gamma}{d\ell}  &  =-\Gamma~\text{,} & \frac{dh}{d\ell}  &
=-h~\text{.}%
\end{align}
Now $h$ became RG-irrelevant. However, the ratio $h/\Gamma$ is
exactly marginal and parametrizes the new finite-field
Fermi-liquid fixed points.

\bibliography{astrings,kondo2007}

\begin{thebibliography}{62}
\expandafter\ifx\csname natexlab\endcsname\relax\def\natexlab#1{#1}\fi
\expandafter\ifx\csname bibnamefont\endcsname\relax
  \def\bibnamefont#1{#1}\fi
\expandafter\ifx\csname bibfnamefont\endcsname\relax
  \def\bibfnamefont#1{#1}\fi
\expandafter\ifx\csname citenamefont\endcsname\relax
  \def\citenamefont#1{#1}\fi
\expandafter\ifx\csname url\endcsname\relax
  \def\url#1{\texttt{#1}}\fi
\expandafter\ifx\csname urlprefix\endcsname\relax\def\urlprefix{URL }\fi
\providecommand{\bibinfo}[2]{#2}
\providecommand{\eprint}[2][]{\url{#2}}

\bibitem[{\citenamefont{Cox}(1987)}]{cox1987}
\bibinfo{author}{\bibfnamefont{D.~L.} \bibnamefont{Cox}},
  \bibinfo{journal}{Phys. Rev. Lett.} \textbf{\bibinfo{volume}{59}},
  \bibinfo{pages}{1240} (\bibinfo{year}{1987}), \bibinfo{note}{{\it Erratum:
  ibid.} {\bf 61}, 1527 (1988)}.

\bibitem[{\citenamefont{Nozi\`eres and Blandin}(1980)}]{Nozieres1980}
\bibinfo{author}{\bibfnamefont{P.}~\bibnamefont{Nozi\`eres}} \bibnamefont{and}
  \bibinfo{author}{\bibfnamefont{A.}~\bibnamefont{Blandin}},
  \bibinfo{journal}{J. Phys. (Paris)} \textbf{\bibinfo{volume}{41}},
  \bibinfo{pages}{193} (\bibinfo{year}{1980}).

\bibitem[{\citenamefont{Kawae et~al.}(2006)\citenamefont{Kawae, Kinoshita,
  Nakaie, Tateiwa, Takeda, Suzuki, and Kitai}}]{kawae2006}
\bibinfo{author}{\bibfnamefont{T.}~\bibnamefont{Kawae}},
  \bibinfo{author}{\bibfnamefont{K.}~\bibnamefont{Kinoshita}},
  \bibinfo{author}{\bibfnamefont{Y.}~\bibnamefont{Nakaie}},
  \bibinfo{author}{\bibfnamefont{N.}~\bibnamefont{Tateiwa}},
  \bibinfo{author}{\bibfnamefont{K.}~\bibnamefont{Takeda}},
  \bibinfo{author}{\bibfnamefont{H.~S.} \bibnamefont{Suzuki}},
  \bibnamefont{and} \bibinfo{author}{\bibfnamefont{T.}~\bibnamefont{Kitai}},
  \bibinfo{journal}{Phys. Rev. Lett.} \textbf{\bibinfo{volume}{96}},
  \bibinfo{pages}{027210} (\bibinfo{year}{2006}).

\bibitem[{\citenamefont{Cox and Zawadowski}(1998)}]{cox1998}
\bibinfo{author}{\bibfnamefont{D.~L.~A.} \bibnamefont{Cox}} \bibnamefont{and}
  \bibinfo{author}{\bibfnamefont{A.}~\bibnamefont{Zawadowski}},
  \bibinfo{journal}{Adv. Phys.} \textbf{\bibinfo{volume}{47}},
  \bibinfo{pages}{599} (\bibinfo{year}{1998}).

\bibitem[{\citenamefont{Hotta}(2005)}]{hotta2005}
\bibinfo{author}{\bibfnamefont{T.}~\bibnamefont{Hotta}},
  \bibinfo{journal}{Phys. Rev. Lett.} \textbf{\bibinfo{volume}{94}},
  \bibinfo{pages}{067003} (\bibinfo{year}{2005}).

\bibitem[{\citenamefont{Hotta}(2006)}]{hotta2006}
\bibinfo{author}{\bibfnamefont{T.}~\bibnamefont{Hotta}},
  \bibinfo{journal}{Phys. Rev. Lett.} \textbf{\bibinfo{volume}{96}},
  \bibinfo{pages}{197201} (\bibinfo{year}{2006}).

\bibitem[{\citenamefont{Ramirez et~al.}(1994)\citenamefont{Ramirez, Chandra,
  Coleman, Fisk, Smith, and Ott}}]{ramirez1994}
\bibinfo{author}{\bibfnamefont{A.~P.} \bibnamefont{Ramirez}},
  \bibinfo{author}{\bibfnamefont{P.}~\bibnamefont{Chandra}},
  \bibinfo{author}{\bibfnamefont{P.}~\bibnamefont{Coleman}},
  \bibinfo{author}{\bibfnamefont{Z.}~\bibnamefont{Fisk}},
  \bibinfo{author}{\bibfnamefont{J.~L.} \bibnamefont{Smith}}, \bibnamefont{and}
  \bibinfo{author}{\bibfnamefont{H.~R.} \bibnamefont{Ott}},
  \bibinfo{journal}{Phys. Rev. Lett.} \textbf{\bibinfo{volume}{73}},
  \bibinfo{pages}{3018} (\bibinfo{year}{1994}).

\bibitem[{\citenamefont{Aliev et~al.}(1995)\citenamefont{Aliev, Mfarrej,
  Vieira, Villar, and Martinez}}]{aliev1995}
\bibinfo{author}{\bibfnamefont{F.~G.} \bibnamefont{Aliev}},
  \bibinfo{author}{\bibfnamefont{H.~E.} \bibnamefont{Mfarrej}},
  \bibinfo{author}{\bibfnamefont{S.}~\bibnamefont{Vieira}},
  \bibinfo{author}{\bibfnamefont{R.}~\bibnamefont{Villar}}, \bibnamefont{and}
  \bibinfo{author}{\bibfnamefont{J.~L.} \bibnamefont{Martinez}},
  \bibinfo{journal}{Europhys. Lett.} \textbf{\bibinfo{volume}{32}},
  \bibinfo{pages}{765} (\bibinfo{year}{1995}).

\bibitem[{\citenamefont{Harrison et~al.}(2001)\citenamefont{Harrison, Balicas,
  Teklu, Goodrich, Brooks, Cooley, and Smith}}]{harrison2001}
\bibinfo{author}{\bibfnamefont{N.}~\bibnamefont{Harrison}},
  \bibinfo{author}{\bibfnamefont{L.}~\bibnamefont{Balicas}},
  \bibinfo{author}{\bibfnamefont{A.~A.} \bibnamefont{Teklu}},
  \bibinfo{author}{\bibfnamefont{R.~G.} \bibnamefont{Goodrich}},
  \bibinfo{author}{\bibfnamefont{J.~S.} \bibnamefont{Brooks}},
  \bibinfo{author}{\bibfnamefont{J.~C.} \bibnamefont{Cooley}},
  \bibnamefont{and} \bibinfo{author}{\bibfnamefont{J.~L.} \bibnamefont{Smith}},
  \bibinfo{journal}{Phys. Rev. B} \textbf{\bibinfo{volume}{63}},
  \bibinfo{pages}{081101(R)} (\bibinfo{year}{2001}).

\bibitem[{\citenamefont{Schiller et~al.}(1998)\citenamefont{Schiller, Anders,
  and Cox}}]{schiller1998}
\bibinfo{author}{\bibfnamefont{A.}~\bibnamefont{Schiller}},
  \bibinfo{author}{\bibfnamefont{F.~B.} \bibnamefont{Anders}},
  \bibnamefont{and} \bibinfo{author}{\bibfnamefont{D.~L.} \bibnamefont{Cox}},
  \bibinfo{journal}{Phys. Rev. Lett.} \textbf{\bibinfo{volume}{81}},
  \bibinfo{pages}{3235} (\bibinfo{year}{1998}).

\bibitem[{\citenamefont{Koga and Cox}(1999)}]{koga1999}
\bibinfo{author}{\bibfnamefont{M.}~\bibnamefont{Koga}} \bibnamefont{and}
  \bibinfo{author}{\bibfnamefont{D.~L.} \bibnamefont{Cox}},
  \bibinfo{journal}{Phys. Rev. Lett.} \textbf{\bibinfo{volume}{82}},
  \bibinfo{pages}{2575} (\bibinfo{year}{1999}).

\bibitem[{\citenamefont{Bolech and Andrei}(2002)}]{bolech2002}
\bibinfo{author}{\bibfnamefont{C.~J.} \bibnamefont{Bolech}} \bibnamefont{and}
  \bibinfo{author}{\bibfnamefont{N.}~\bibnamefont{Andrei}},
  \bibinfo{journal}{Phys. Rev. Lett.} \textbf{\bibinfo{volume}{88}},
  \bibinfo{pages}{237206} (\bibinfo{year}{2002}).

\bibitem[{\citenamefont{Onimaru et~al.}(2005)\citenamefont{Onimaru, Sakakibara,
  Aso, Yoshizawa, Suzuki, and Takeuchi}}]{onimaru2005}
\bibinfo{author}{\bibfnamefont{T.}~\bibnamefont{Onimaru}},
  \bibinfo{author}{\bibfnamefont{T.}~\bibnamefont{Sakakibara}},
  \bibinfo{author}{\bibfnamefont{N.}~\bibnamefont{Aso}},
  \bibinfo{author}{\bibfnamefont{H.}~\bibnamefont{Yoshizawa}},
  \bibinfo{author}{\bibfnamefont{H.~S.} \bibnamefont{Suzuki}},
  \bibnamefont{and} \bibinfo{author}{\bibfnamefont{T.}~\bibnamefont{Takeuchi}},
  \bibinfo{journal}{Phys. Rev. Lett.} \textbf{\bibinfo{volume}{94}},
  \bibinfo{pages}{197201} (\bibinfo{year}{2005}).

\bibitem[{\citenamefont{Bauer et~al.}(2002)\citenamefont{Bauer, Frederick, Ho,
  Zapf, and Maple}}]{bauer2002}
\bibinfo{author}{\bibfnamefont{E.~D.} \bibnamefont{Bauer}},
  \bibinfo{author}{\bibfnamefont{N.~A.} \bibnamefont{Frederick}},
  \bibinfo{author}{\bibfnamefont{P.-C.} \bibnamefont{Ho}},
  \bibinfo{author}{\bibfnamefont{V.~S.} \bibnamefont{Zapf}}, \bibnamefont{and}
  \bibinfo{author}{\bibfnamefont{M.~B.} \bibnamefont{Maple}},
  \bibinfo{journal}{Phys. Rev. B} \textbf{\bibinfo{volume}{65}},
  \bibinfo{pages}{100506(R)} (\bibinfo{year}{2002}).

\bibitem[{\citenamefont{Bauer et~al.}(2006)\citenamefont{Bauer, Ho, Maple,
  Schauerte, Cox, and Anders}}]{bauer2006}
\bibinfo{author}{\bibfnamefont{E.~D.} \bibnamefont{Bauer}},
  \bibinfo{author}{\bibfnamefont{P.-C.} \bibnamefont{Ho}},
  \bibinfo{author}{\bibfnamefont{M.~B.} \bibnamefont{Maple}},
  \bibinfo{author}{\bibfnamefont{T.}~\bibnamefont{Schauerte}},
  \bibinfo{author}{\bibfnamefont{D.~L.} \bibnamefont{Cox}}, \bibnamefont{and}
  \bibinfo{author}{\bibfnamefont{F.~B.} \bibnamefont{Anders}},
  \bibinfo{journal}{Phys. Rev. B} \textbf{\bibinfo{volume}{73}},
  \bibinfo{pages}{094511} (\bibinfo{year}{2006}).

\bibitem[{\citenamefont{Maple}(2005)}]{maple2005}
\bibinfo{author}{\bibfnamefont{M.~B.} \bibnamefont{Maple}},
  \bibinfo{journal}{J. Phys. Soc. Jpn.} \textbf{\bibinfo{volume}{74}},
  \bibinfo{pages}{222} (\bibinfo{year}{2005}).

\bibitem[{\citenamefont{Berman et~al.}(1999)\citenamefont{Berman, Zhitenev,
  Ashoori, and Shayegan}}]{berman1999}
\bibinfo{author}{\bibfnamefont{D.}~\bibnamefont{Berman}},
  \bibinfo{author}{\bibfnamefont{N.~B.} \bibnamefont{Zhitenev}},
  \bibinfo{author}{\bibfnamefont{R.~C.} \bibnamefont{Ashoori}},
  \bibnamefont{and} \bibinfo{author}{\bibfnamefont{M.}~\bibnamefont{Shayegan}},
  \bibinfo{journal}{Phys. Rev. Lett.} \textbf{\bibinfo{volume}{82}},
  \bibinfo{pages}{161} (\bibinfo{year}{1999}).

\bibitem[{\citenamefont{Oreg and Goldhaber-Gordon}(2003)}]{oreg2003}
\bibinfo{author}{\bibfnamefont{Y.}~\bibnamefont{Oreg}} \bibnamefont{and}
  \bibinfo{author}{\bibfnamefont{D.}~\bibnamefont{Goldhaber-Gordon}},
  \bibinfo{journal}{Phys. Rev. Lett.} \textbf{\bibinfo{volume}{90}},
  \bibinfo{pages}{136602} (\bibinfo{year}{2003}).

\bibitem[{\citenamefont{Shah and Millis}(2003)}]{shah2003}
\bibinfo{author}{\bibfnamefont{N.}~\bibnamefont{Shah}} \bibnamefont{and}
  \bibinfo{author}{\bibfnamefont{A.~J.} \bibnamefont{Millis}},
  \bibinfo{journal}{Phys. Rev. Lett.} \textbf{\bibinfo{volume}{91}},
  \bibinfo{pages}{147204} (\bibinfo{year}{2003}).

\bibitem[{\citenamefont{Kakashvili and Johannesson}(2007)}]{kakashvili2007}
\bibinfo{author}{\bibfnamefont{P.}~\bibnamefont{Kakashvili}} \bibnamefont{and}
  \bibinfo{author}{\bibfnamefont{H.}~\bibnamefont{Johannesson}},
  \bibinfo{journal}{Europhys. Lett.} \textbf{\bibinfo{volume}{79}},
  \bibinfo{pages}{47004} (\bibinfo{year}{2007}).

\bibitem[{\citenamefont{Potok et~al.}(2007)\citenamefont{Potok, Rau, Shtrikman,
  Oreg, and Goldhaber-Gordon}}]{potok2007}
\bibinfo{author}{\bibfnamefont{R.~M.} \bibnamefont{Potok}},
  \bibinfo{author}{\bibfnamefont{I.~G.} \bibnamefont{Rau}},
  \bibinfo{author}{\bibfnamefont{H.}~\bibnamefont{Shtrikman}},
  \bibinfo{author}{\bibfnamefont{Y.}~\bibnamefont{Oreg}}, \bibnamefont{and}
  \bibinfo{author}{\bibfnamefont{D.}~\bibnamefont{Goldhaber-Gordon}},
  \bibinfo{journal}{Nature (London)} \textbf{\bibinfo{volume}{446}},
  \bibinfo{pages}{167} (\bibinfo{year}{2007}).

\bibitem[{\citenamefont{Bolech and Shah}(2005)}]{bolech2005b}
\bibinfo{author}{\bibfnamefont{C.~J.} \bibnamefont{Bolech}} \bibnamefont{and}
  \bibinfo{author}{\bibfnamefont{N.}~\bibnamefont{Shah}},
  \bibinfo{journal}{Phys. Rev. Lett.} \textbf{\bibinfo{volume}{95}},
  \bibinfo{pages}{036801} (\bibinfo{year}{2005}).

\bibitem[{\citenamefont{Kondo}(1964)}]{kondo1964}
\bibinfo{author}{\bibfnamefont{J.}~\bibnamefont{Kondo}},
  \bibinfo{journal}{Prog. Theor. Phys.} \textbf{\bibinfo{volume}{32}},
  \bibinfo{pages}{37} (\bibinfo{year}{1964}).

\bibitem[{\citenamefont{Wilson}(1975)}]{wilson1975}
\bibinfo{author}{\bibfnamefont{K.~G.} \bibnamefont{Wilson}},
  \bibinfo{journal}{Rev. Mod. Phys.} \textbf{\bibinfo{volume}{47}},
  \bibinfo{pages}{773} (\bibinfo{year}{1975}).

\bibitem[{\citenamefont{Andrei}(1980)}]{andrei1980}
\bibinfo{author}{\bibfnamefont{N.}~\bibnamefont{Andrei}},
  \bibinfo{journal}{Phys. Rev. Lett.} \textbf{\bibinfo{volume}{45}},
  \bibinfo{pages}{379} (\bibinfo{year}{1980}).

\bibitem[{\citenamefont{Wiegmann}(1980)}]{wiegmann1980}
\bibinfo{author}{\bibfnamefont{P.~B.} \bibnamefont{Wiegmann}},
  \bibinfo{journal}{Phys. Lett. A} \textbf{\bibinfo{volume}{80}},
  \bibinfo{pages}{163} (\bibinfo{year}{1980}).

\bibitem[{\citenamefont{Toulouse}(1969)}]{toulouse1969}
\bibinfo{author}{\bibfnamefont{G.}~\bibnamefont{Toulouse}},
  \bibinfo{journal}{C. R. Acad. Sci.} \textbf{\bibinfo{volume}{268}},
  \bibinfo{pages}{1200} (\bibinfo{year}{1969}).

\bibitem[{\citenamefont{Schlottmann}(1978)}]{schlottmann1978}
\bibinfo{author}{\bibfnamefont{P.}~\bibnamefont{Schlottmann}},
  \bibinfo{journal}{J. Phys. (Paris)} \textbf{\bibinfo{volume}{39}},
  \bibinfo{pages}{1486} (\bibinfo{year}{1978}).

\bibitem[{\citenamefont{Vigman and {Finkel'shte{\v{\i}}n}}(1978)}]{vigman1978}
\bibinfo{author}{\bibfnamefont{P.~B.} \bibnamefont{Vigman}} \bibnamefont{and}
  \bibinfo{author}{\bibfnamefont{A.~M.} \bibnamefont{{Finkel'shte{\v{\i}}n}}},
  \bibinfo{journal}{Sov. Phys. JETP} \textbf{\bibinfo{volume}{48}},
  \bibinfo{pages}{102} (\bibinfo{year}{1978}).

\bibitem[{\citenamefont{Hewson}(1993)}]{Hewson}
\bibinfo{author}{\bibfnamefont{A.~C.} \bibnamefont{Hewson}},
  \emph{\bibinfo{title}{The {K}ondo Problem to {H}eavy {F}ermions}}
  (\bibinfo{publisher}{Cambridge University Press},
  \bibinfo{address}{Cambridge, England}, \bibinfo{year}{1993}).

\bibitem[{\citenamefont{Kotliar and Si}(1996)}]{kotliar1996}
\bibinfo{author}{\bibfnamefont{G.}~\bibnamefont{Kotliar}} \bibnamefont{and}
  \bibinfo{author}{\bibfnamefont{Q.}~\bibnamefont{Si}}, \bibinfo{journal}{Phys.
  Rev. B} \textbf{\bibinfo{volume}{53}}, \bibinfo{pages}{12373}
  (\bibinfo{year}{1996}).

\bibitem[{\citenamefont{Bolech and Iucci}(2006{\natexlab{a}})}]{bolech2006a}
\bibinfo{author}{\bibfnamefont{C.~J.} \bibnamefont{Bolech}} \bibnamefont{and}
  \bibinfo{author}{\bibfnamefont{A.}~\bibnamefont{Iucci}},
  \bibinfo{journal}{Phys. Rev. Lett.} \textbf{\bibinfo{volume}{96}},
  \bibinfo{pages}{056402} (\bibinfo{year}{2006}{\natexlab{a}}).

\bibitem[{\citenamefont{Andrei et~al.}(1983)\citenamefont{Andrei, Furuya, and
  Lowenstein}}]{afl1983}
\bibinfo{author}{\bibfnamefont{N.}~\bibnamefont{Andrei}},
  \bibinfo{author}{\bibfnamefont{K.}~\bibnamefont{Furuya}}, \bibnamefont{and}
  \bibinfo{author}{\bibfnamefont{J.~H.} \bibnamefont{Lowenstein}},
  \bibinfo{journal}{Rev. Mod. Phys.} \textbf{\bibinfo{volume}{55}},
  \bibinfo{pages}{331} (\bibinfo{year}{1983}).

\bibitem[{\citenamefont{Affleck}(1995)}]{affleck1995}
\bibinfo{author}{\bibfnamefont{I.}~\bibnamefont{Affleck}},
  \bibinfo{journal}{Acta Phys. Pol. B} \textbf{\bibinfo{volume}{26}},
  \bibinfo{pages}{1869} (\bibinfo{year}{1995}), \bibinfo{note}{lecture given at
  the {\rm XXXV}th Cracow School of Theoretical Physics, Zakopane, Poland, June
  4$^\text{th}$-14$^\text{th}$ 1995}.

\bibitem[{\citenamefont{Giamarchi}(2004)}]{Giamarchi}
\bibinfo{author}{\bibfnamefont{T.}~\bibnamefont{Giamarchi}},
  \emph{\bibinfo{title}{Quantum Physics in One Dimension}}
  (\bibinfo{publisher}{Clarendon Press}, \bibinfo{address}{Oxford},
  \bibinfo{year}{2004}).

\bibitem[{\citenamefont{Emery and Kivelson}(1992)}]{emery1992}
\bibinfo{author}{\bibfnamefont{V.~J.} \bibnamefont{Emery}} \bibnamefont{and}
  \bibinfo{author}{\bibfnamefont{S.}~\bibnamefont{Kivelson}},
  \bibinfo{journal}{Phys. Rev. B} \textbf{\bibinfo{volume}{46}},
  \bibinfo{pages}{10812} (\bibinfo{year}{1992}).

\bibitem[{\citenamefont{Fabrizio et~al.}(1995)\citenamefont{Fabrizio, Gogolin,
  and Nozi{\`e}res}}]{fabrizio1995}
\bibinfo{author}{\bibfnamefont{M.}~\bibnamefont{Fabrizio}},
  \bibinfo{author}{\bibfnamefont{A.~O.} \bibnamefont{Gogolin}},
  \bibnamefont{and}
  \bibinfo{author}{\bibfnamefont{P.}~\bibnamefont{Nozi{\`e}res}},
  \bibinfo{journal}{Phys. Rev. B} \textbf{\bibinfo{volume}{51}},
  \bibinfo{pages}{16088} (\bibinfo{year}{1995}).

\bibitem[{\citenamefont{Schofield}(1997)}]{schofield1997}
\bibinfo{author}{\bibfnamefont{A.~J.} \bibnamefont{Schofield}},
  \bibinfo{journal}{Phys. Rev. B} \textbf{\bibinfo{volume}{55}},
  \bibinfo{pages}{5627} (\bibinfo{year}{1997}).

\bibitem[{\citenamefont{Ye}(1997)}]{ye1997}
\bibinfo{author}{\bibfnamefont{J.}~\bibnamefont{Ye}}, \bibinfo{journal}{Phys.
  Rev. B} \textbf{\bibinfo{volume}{56}}, \bibinfo{pages}{R489}
  (\bibinfo{year}{1997}).

\bibitem[{\citenamefont{{von Delft} et~al.}(1998)\citenamefont{{von Delft},
  Zar{\'a}nd, and Fabrizio}}]{vondelft1998}
\bibinfo{author}{\bibfnamefont{J.}~\bibnamefont{{von Delft}}},
  \bibinfo{author}{\bibfnamefont{G.}~\bibnamefont{Zar{\'a}nd}},
  \bibnamefont{and} \bibinfo{author}{\bibfnamefont{M.}~\bibnamefont{Fabrizio}},
  \bibinfo{journal}{Phys. Rev. Lett.} \textbf{\bibinfo{volume}{81}},
  \bibinfo{pages}{196} (\bibinfo{year}{1998}).

\bibitem[{\citenamefont{Johannesson et~al.}(2003)\citenamefont{Johannesson,
  Andrei, and Bolech}}]{johannesson2003}
\bibinfo{author}{\bibfnamefont{H.}~\bibnamefont{Johannesson}},
  \bibinfo{author}{\bibfnamefont{N.}~\bibnamefont{Andrei}}, \bibnamefont{and}
  \bibinfo{author}{\bibfnamefont{C.~J.} \bibnamefont{Bolech}},
  \bibinfo{journal}{Phys. Rev. B} \textbf{\bibinfo{volume}{68}},
  \bibinfo{pages}{075112} (\bibinfo{year}{2003}).

\bibitem[{\citenamefont{Johannesson et~al.}(2005)\citenamefont{Johannesson,
  Bolech, and Andrei}}]{johannesson2005}
\bibinfo{author}{\bibfnamefont{H.}~\bibnamefont{Johannesson}},
  \bibinfo{author}{\bibfnamefont{C.~J.} \bibnamefont{Bolech}},
  \bibnamefont{and} \bibinfo{author}{\bibfnamefont{N.}~\bibnamefont{Andrei}},
  \bibinfo{journal}{Phys. Rev. B} \textbf{\bibinfo{volume}{71}},
  \bibinfo{pages}{195107} (\bibinfo{year}{2005}).

\bibitem[{\citenamefont{Anders}(2005)}]{anders2005}
\bibinfo{author}{\bibfnamefont{F.~B.} \bibnamefont{Anders}},
  \bibinfo{journal}{Phys. Rev. B} \textbf{\bibinfo{volume}{71}},
  \bibinfo{pages}{121101(R)} (\bibinfo{year}{2005}).

\bibitem[{\citenamefont{Bolech and Andrei}(2005)}]{bolech2005a}
\bibinfo{author}{\bibfnamefont{C.~J.} \bibnamefont{Bolech}} \bibnamefont{and}
  \bibinfo{author}{\bibfnamefont{N.}~\bibnamefont{Andrei}},
  \bibinfo{journal}{Phys. Rev. B} \textbf{\bibinfo{volume}{71}},
  \bibinfo{pages}{205104} (\bibinfo{year}{2005}).

\bibitem[{\citenamefont{Haldane}(1981)}]{haldane1981}
\bibinfo{author}{\bibfnamefont{F.~D.~M.} \bibnamefont{Haldane}},
  \bibinfo{journal}{J. Phys. C} \textbf{\bibinfo{volume}{14}},
  \bibinfo{pages}{2585} (\bibinfo{year}{1981}).

\bibitem[{\citenamefont{{von Delft} and Schoeller}(1998)}]{vondelft1998a}
\bibinfo{author}{\bibfnamefont{J.}~\bibnamefont{{von Delft}}} \bibnamefont{and}
  \bibinfo{author}{\bibfnamefont{H.}~\bibnamefont{Schoeller}},
  \bibinfo{journal}{Annalen der Phys.} \textbf{\bibinfo{volume}{7}},
  \bibinfo{pages}{225} (\bibinfo{year}{1998}).

\bibitem[{\citenamefont{Gogolin et~al.}(1998)\citenamefont{Gogolin, Nersesyan,
  and Tsvelik}}]{Gogolin}
\bibinfo{author}{\bibfnamefont{A.~O.} \bibnamefont{Gogolin}},
  \bibinfo{author}{\bibfnamefont{A.~A.} \bibnamefont{Nersesyan}},
  \bibnamefont{and} \bibinfo{author}{\bibfnamefont{A.~M.}
  \bibnamefont{Tsvelik}}, \emph{\bibinfo{title}{Bosonization and Strongly
  Correlated Systems}} (\bibinfo{publisher}{Cambridge University Press},
  \bibinfo{address}{Cambridge, England}, \bibinfo{year}{1998}).

\bibitem[{\citenamefont{Zar{\'a}nd and {von Delft}}(2000)}]{zarand2000}
\bibinfo{author}{\bibfnamefont{G.}~\bibnamefont{Zar{\'a}nd}} \bibnamefont{and}
  \bibinfo{author}{\bibfnamefont{J.}~\bibnamefont{{von Delft}}},
  \bibinfo{journal}{Phys. Rev. B} \textbf{\bibinfo{volume}{61}},
  \bibinfo{pages}{6918} (\bibinfo{year}{2000}).

\bibitem[{\citenamefont{Clarke et~al.}(1993)\citenamefont{Clarke, Giamarchi,
  and Shraiman}}]{clarke1993}
\bibinfo{author}{\bibfnamefont{D.~G.} \bibnamefont{Clarke}},
  \bibinfo{author}{\bibfnamefont{T.}~\bibnamefont{Giamarchi}},
  \bibnamefont{and} \bibinfo{author}{\bibfnamefont{B.~I.}
  \bibnamefont{Shraiman}}, \bibinfo{journal}{Phys. Rev. B}
  \textbf{\bibinfo{volume}{48}}, \bibinfo{pages}{7070} (\bibinfo{year}{1993}).

\bibitem[{\citenamefont{Sengupta and Georges}(1994)}]{sengupta1994}
\bibinfo{author}{\bibfnamefont{A.~M.} \bibnamefont{Sengupta}} \bibnamefont{and}
  \bibinfo{author}{\bibfnamefont{A.}~\bibnamefont{Georges}},
  \bibinfo{journal}{Phys. Rev. B} \textbf{\bibinfo{volume}{49}},
  \bibinfo{pages}{R10020} (\bibinfo{year}{1994}).

\bibitem[{\citenamefont{Maldacena and Ludwig}(1997)}]{maldacena1997}
\bibinfo{author}{\bibfnamefont{J.~M.} \bibnamefont{Maldacena}}
  \bibnamefont{and} \bibinfo{author}{\bibfnamefont{A.~W.~W.}
  \bibnamefont{Ludwig}}, \bibinfo{journal}{Nucl. Phys. B}
  \textbf{\bibinfo{volume}{506}}, \bibinfo{pages}{565} (\bibinfo{year}{1997}).

\bibitem[{\citenamefont{Ye}(1998)}]{ye1998}
\bibinfo{author}{\bibfnamefont{J.}~\bibnamefont{Ye}}, \bibinfo{journal}{Nucl.
  Phys. B} \textbf{\bibinfo{volume}{512}}, \bibinfo{pages}{543}
  (\bibinfo{year}{1998}).

\bibitem[{\citenamefont{Shankar}(1994)}]{shankar1994}
\bibinfo{author}{\bibfnamefont{R.}~\bibnamefont{Shankar}},
  \bibinfo{journal}{Rev. Mod. Phys.} \textbf{\bibinfo{volume}{66}},
  \bibinfo{pages}{129} (\bibinfo{year}{1994}).

\bibitem[{\citenamefont{Moustakas and Fisher}(1996)}]{moustakas1996}
\bibinfo{author}{\bibfnamefont{A.~L.} \bibnamefont{Moustakas}}
  \bibnamefont{and} \bibinfo{author}{\bibfnamefont{D.~S.}
  \bibnamefont{Fisher}}, \bibinfo{journal}{Phys. Rev. B}
  \textbf{\bibinfo{volume}{53}}, \bibinfo{pages}{4300} (\bibinfo{year}{1996}).

\bibitem[{\citenamefont{Gan}(1995)}]{gan1995}
\bibinfo{author}{\bibfnamefont{J.}~\bibnamefont{Gan}}, \bibinfo{journal}{Phys.
  Rev. B} \textbf{\bibinfo{volume}{51}}, \bibinfo{pages}{8287}
  (\bibinfo{year}{1995}).

\bibitem[{\citenamefont{Tsvelik and Wiegmann}(1983)}]{tsvelik1983}
\bibinfo{author}{\bibfnamefont{A.~M.} \bibnamefont{Tsvelik}} \bibnamefont{and}
  \bibinfo{author}{\bibfnamefont{P.~B.} \bibnamefont{Wiegmann}},
  \bibinfo{journal}{Adv. Phys.} \textbf{\bibinfo{volume}{32}},
  \bibinfo{pages}{453} (\bibinfo{year}{1983}).

\bibitem[{\citenamefont{Zar{\'a}nd et~al.}(2002)\citenamefont{Zar{\'a}nd,
  Costi, Jerez, and Andrei}}]{zarand2002}
\bibinfo{author}{\bibfnamefont{G.}~\bibnamefont{Zar{\'a}nd}},
  \bibinfo{author}{\bibfnamefont{T.}~\bibnamefont{Costi}},
  \bibinfo{author}{\bibfnamefont{A.}~\bibnamefont{Jerez}}, \bibnamefont{and}
  \bibinfo{author}{\bibfnamefont{N.}~\bibnamefont{Andrei}},
  \bibinfo{journal}{Phys. Rev. B} \textbf{\bibinfo{volume}{65}},
  \bibinfo{pages}{134416} (\bibinfo{year}{2002}).

\bibitem[{\citenamefont{Bolech and Iucci}(2006{\natexlab{b}})}]{bolech2006b}
\bibinfo{author}{\bibfnamefont{C.~J.} \bibnamefont{Bolech}} \bibnamefont{and}
  \bibinfo{author}{\bibfnamefont{A.}~\bibnamefont{Iucci}},
  \bibinfo{journal}{Physica B} \textbf{\bibinfo{volume}{378-80}},
  \bibinfo{pages}{171} (\bibinfo{year}{2006}{\natexlab{b}}).

\bibitem[{\citenamefont{Tsvelik}(1995)}]{tsvelik1995}
\bibinfo{author}{\bibfnamefont{A.~M.} \bibnamefont{Tsvelik}},
  \bibinfo{journal}{Phys. Rev. B} \textbf{\bibinfo{volume}{52}},
  \bibinfo{pages}{4366} (\bibinfo{year}{1995}).

\bibitem[{\citenamefont{Ferrero et~al.}(2007)\citenamefont{Ferrero, {De Leo},
  Lecheminant, and Fabrizio}}]{ferrero2007}
\bibinfo{author}{\bibfnamefont{M.}~\bibnamefont{Ferrero}},
  \bibinfo{author}{\bibfnamefont{L.}~\bibnamefont{{De Leo}}},
  \bibinfo{author}{\bibfnamefont{P.}~\bibnamefont{Lecheminant}},
  \bibnamefont{and} \bibinfo{author}{\bibfnamefont{M.}~\bibnamefont{Fabrizio}},
  \bibinfo{journal}{J. Phys.: Condens. Matter} \textbf{\bibinfo{volume}{19}},
  \bibinfo{pages}{433201} (\bibinfo{year}{2007}).

\bibitem[{\citenamefont{Schauerte et~al.}(2005)\citenamefont{Schauerte, Cox,
  Noack, {van Dongen}, and Batista}}]{schauerte2005}
\bibinfo{author}{\bibfnamefont{T.}~\bibnamefont{Schauerte}},
  \bibinfo{author}{\bibfnamefont{D.~L.} \bibnamefont{Cox}},
  \bibinfo{author}{\bibfnamefont{R.~M.} \bibnamefont{Noack}},
  \bibinfo{author}{\bibfnamefont{P.~G.~J.} \bibnamefont{{van Dongen}}},
  \bibnamefont{and} \bibinfo{author}{\bibfnamefont{C.~D.}
  \bibnamefont{Batista}}, \bibinfo{journal}{Phys. Rev. Lett.}
  \textbf{\bibinfo{volume}{94}}, \bibinfo{pages}{147201}
  (\bibinfo{year}{2005}).

\bibitem[{\citenamefont{Dickey et~al.}(1997)\citenamefont{Dickey, {de Andrade},
  Herrmann, Maple, Aliev, and Villar}}]{dickey1997}
\bibinfo{author}{\bibfnamefont{R.~P.} \bibnamefont{Dickey}},
  \bibinfo{author}{\bibfnamefont{M.~C.} \bibnamefont{{de Andrade}}},
  \bibinfo{author}{\bibfnamefont{J.}~\bibnamefont{Herrmann}},
  \bibinfo{author}{\bibfnamefont{M.~B.} \bibnamefont{Maple}},
  \bibinfo{author}{\bibfnamefont{F.~G.} \bibnamefont{Aliev}}, \bibnamefont{and}
  \bibinfo{author}{\bibfnamefont{R.}~\bibnamefont{Villar}},
  \bibinfo{journal}{Phys. Rev. B} \textbf{\bibinfo{volume}{56}},
  \bibinfo{pages}{11169} (\bibinfo{year}{1997}).

\end{thebibliography}

\end{subequations}
\end{document}